\documentclass{egpubl}
\usepackage{pg2025}

\SpecialIssuePaper         

\CGFStandardLicense

\usepackage[T1]{fontenc}
\usepackage{dfadobe}  
\usepackage{booktabs}
\usepackage{cite}  
\usepackage{makecell}
\BibtexOrBiblatex
\electronicVersion
\PrintedOrElectronic
\ifpdf \usepackage[pdftex]{graphicx} \pdfcompresslevel=9
\else \usepackage[dvips]{graphicx} \fi

\usepackage{egweblnk} 

\usepackage{multirow}
\usepackage{enumitem}
\usepackage{array}
\usepackage{tabularx}
\usepackage{wrapfig}
\usepackage{graphicx}
\usepackage{amsmath}
\usepackage{amsfonts}

\title[StyleMM]%
      {StyleMM: Stylized 3D Morphable Face Model via \\
Text-Driven Aligned Image Translation}

\author[Seungmi Lee, Kwan Yun, and Junyong Noh]
{\parbox{\textwidth}{\centering Seungmi Lee$^{*}$ \orcid{0009-0007-4897-6271}\quad\quad 
Kwan Yun$^{*}$ \orcid{0000-0002-0205-5203}\quad\quad 
Junyong Noh \orcid{0000-0003-1925-3326}}
        \\
{\parbox{\textwidth}{\centering KAIST, Visual Media Lab\\
         $^*$Contributed equally to this work
       }
}
}

\begin{document}

\newcommand{\fixq}[1]{\textcolor{red}{#1}}

\maketitle
\begin{abstract}
    
We introduce StyleMM, a novel framework that can construct a stylized 3D Morphable Model (3DMM) based on user-defined text descriptions specifying a target style.
Building upon a pre-trained mesh deformation network and a texture generator for original 3DMM-based realistic human faces, our approach fine-tunes these models using stylized facial images generated via text-guided image-to-image (i2i) translation with a diffusion model, which serve as stylization targets for the rendered mesh. To prevent undesired changes in identity, facial alignment, or expressions during i2i translation, we introduce a stylization method that explicitly preserves the facial attributes of the source image. By maintaining these critical attributes during image stylization, the proposed approach ensures consistent 3D style transfer across the 3DMM parameter space through image-based training. Once trained, StyleMM enables feed-forward generation of stylized face meshes with explicit control over shape, expression, and texture parameters, producing meshes with consistent vertex connectivity and animatability. Quantitative and qualitative evaluations demonstrate that our approach outperforms state-of-the-art methods in terms of identity-level facial diversity and stylization capability. The code and videos are available at \url{kwanyun.github.io/stylemm_page}.



\end{abstract}  

\section{Introduction}
A parametric face model allows for the instant creation of diverse 3D avatars by varying adjustable parameters. This significantly reduces the labor and cost associated with individually modeling each avatar for digital content production. The resulting faces share a unified structural framework—including vertex connectivity, UV maps, and rigs—which enables efficient asset management in editing and reuse tasks. Moreover, the interactive generation process through input parameters facilitates seamless expansion into personalized and customized content. A well-known example is the 3D Morphable Face Model (3DMM), which parameterizes both the shape and expression of human faces and is widely employed for head avatar generation.

Beyond realistic human representations, a parametric face model designed for artistic expression or fictional appearances is particularly valuable in applications such as film, animation, and game production. We define such models as stylized 3D morphable face models (stylized 3DMMs). While stylized 3DMMs must capture a broad range of expressive styles beyond those represented in realistic face models, they should also support identity-level facial variation and maintain a consistent underlying structure across generated assets. Furthermore, to enhance user engagement and creative flexibility, the models should offer intuitive control, enabling creators to rapidly explore diverse appearance variations and achieve desired visual outcomes.

To formalize these requirements, we identify three key elements of a stylized 3DMM. The first two are adopted from the definition of realistic 3DMMs~\cite{egger20203d,blanz1999morphable}, while the third is specific to stylized models.
\begin{enumerate}
\item All generated faces share dense point-to-point correspondences, ensuring consistent mesh structure across instances. (Maintained Correspondence)
\item Facial shape and color are disentangled and can be independently controlled. (Disentangled Control)
\item Expressive stylization of both geometry and texture that extends beyond realistic face models. (Stylization Beyond Realistic Geometry and Texture)
\end{enumerate}
Previous studies have successfully addressed one or two of these elements. Unfortunately, no existing method fulfills all three key requirements of a stylized 3DMM. A detailed discussion is provided in the following paragraphs.

\begin{table}[t]
\setlength{\tabcolsep}{2pt}
\renewcommand{\arraystretch}{1.4}
\caption{Comparison of different 3D face stylization methods. Circles indicate that the method fully meets a criterion, while triangles indicate partial fulfillment -for example, when the geometry satisfies the criterion but the texture does not. StyleMM satisfies all three elements of stylized parametric face models.}
\centering
\resizebox{1\linewidth}{!}{
\begin{tabular}{lccc}
\hline
Method & \makecell{Maintained \\ Correspondence} & \makecell{Disentangled \\  Control} & \makecell{Beyond Realistic \\ Geometry \& Texture} \\
\hline
StyleMM (Ours)          & $\bigcirc  $  & $\bigcirc   $ & $ \bigcirc   $ \\
LeGO~\cite{yoon2024lego}          & $\bigcirc  $  & $\triangle  $ & $ \triangle $ \\
CLIPFace~\cite{aneja2022clipface}          & $\bigcirc  $  & $\bigcirc  $ & $ \triangle $ \\
Ultravatar~\cite{zhou2024ultravatar}    & $\bigcirc  $  & $\triangle  $ & $ \triangle $ \\
HeadEvolver~\cite{wang2024headevolver}   & $\bigcirc  $  & $\times $ & $ \bigcirc   $ \\
ToonifyGB\cite{ju2025toonifygb}& $\triangle $ & $\times $ & $ \bigcirc   $ \\\hline 
\end{tabular} \vspace{-3mm}
}\label{tab:components}
\end{table}

Recent neural rendering-based methods~\cite{abdal20233davatargan, kim2023datid,han2023headsculpt,zhou2024headstudio,batuhan2024identity,ju2025toonifygb,song2024texttoon} have demonstrated promising capabilities in generating stylized 3D faces. Because these methods do not use consistent connectivity, {however,} they lack registered mapping and structural coherence, making it difficult to maintain dense point-to-point correspondence across identities. Furthermore, shape and color are entangled in these models due to their representations, which employ NeRF~\cite{mildenhall2021nerf} or 3D Gaussian Splatting~\cite{kerbl20233d}.

Template face deformation methods based on iterative optimization~\cite{liao2024tada, wang2024headevolver, zhou2024ultravatar,zhang2023Dreamface} have also been proposed. These methods generally leverage priors from pre-trained diffusion models to refine both the geometry and texture of a realistic 3DMM. These techniques successfully produce high-quality stylized meshes typically for a single identity from each optimization process. One drawback of these methods is that geometric and textural components are typically entangled without unified texture mapping. Additionally, the per-character optimization required by these methods introduces substantial computational overhead, significantly restricting their scalability for broader applications.

A few studies have extended parametric face models to stylized domains through deep-learning-based surface deformation methods~\cite{jung2022deep, yoon2024lego}. These approaches primarily focus on geometric stylization by deforming a template mesh according to input latent codes, thereby maintaining dense correspondences and enabling controlled geometric deformation. However, these methods do not incorporate texture manipulation, limiting their ability to independently control and stylize facial textures. Moreover, they require stylized 3D datasets for training, restricting accessibility for novice users and constraining the range of achievable stylistic diversity.

In this work, we propose \textit{StyleMM}, an automated framework that fulfills the three essential elements of a stylized 3DMM: dense point-to-point correspondence, separate and independent control over shape and texture through effective disentanglement, and stylization capabilities that extend beyond the limits of realistic face models. As summarized in Table~\ref{tab:components}, StyleMM uniquely addresses these criteria by directly leveraging textual descriptions of desired styles, which eliminates the need for stylized 3D datasets. Specifically, StyleMM employs diffusion models and their text-guided image-to-image (i2i) translation capabilities. It fine-tunes a surface deformation network and a texture generator, initially designed for realistic 3DMM faces, using rendered stylized images as training targets. This approach ensures dense correspondence across generated meshes, independently controllable shape and texture parameters, and the capability for extensive geometric and textural stylization.

\begin{figure}
    \centering
    \includegraphics[width=\linewidth]{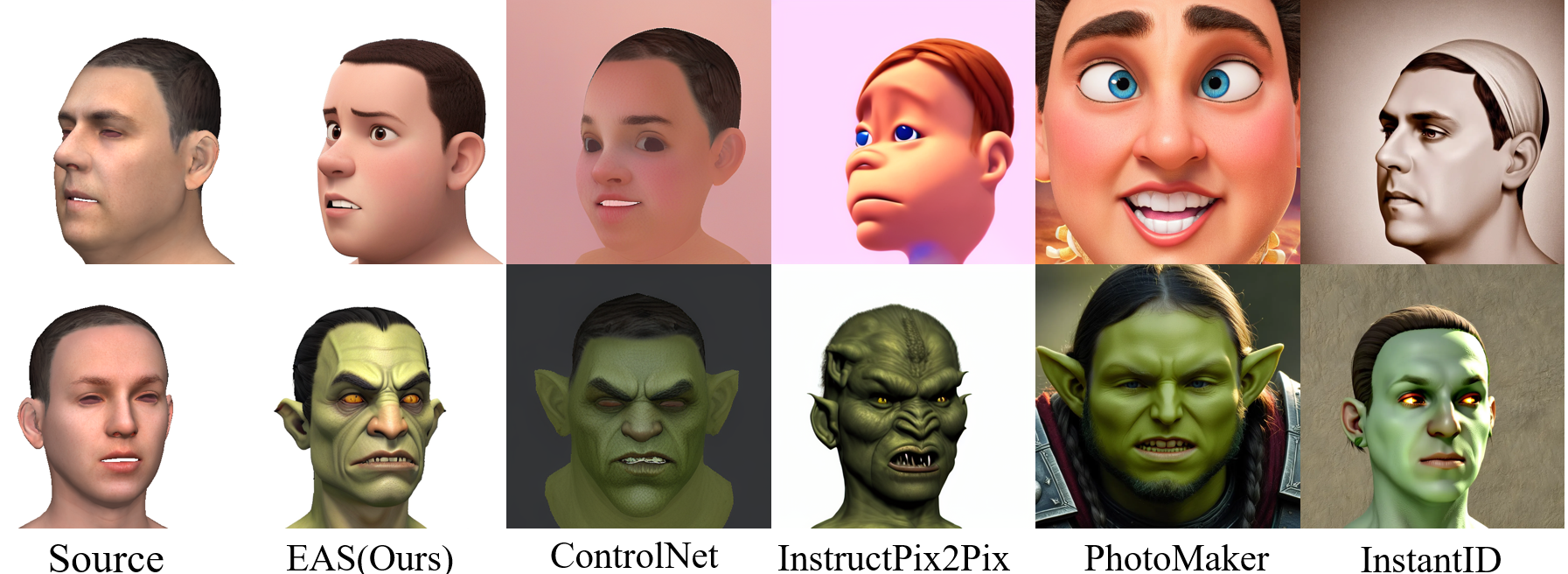}
    \caption{Comparison between EAS and baseline text based i2i methods. Top row: Disney character, Bottom row: green Orc. See Section ~\ref{sec:image-stylization} for experimental details.} \vspace{-2mm}
    \label{fig:i2i_comparison}
\end{figure}

In the process of using stylized images as deformation targets, we found that existing i2i translation methods often alter facial expressions, alignments, or facial structures, posing a significant challenge for training 3D models based on rendered images (See Figure~\ref{fig:i2i_comparison}). Even a small misalignment can hinder the intended deformation of the mesh. To address this, we propose Explicit Attribute-preserving Stylization (EAS), which utilizes explicit facial attributes of sparse facial landmarks, head rotation, and expression. EAS helps preserve key facial attributes, improving the quality and consistency of image-based training.

Building on these images, we train a stylized 3DMM to model various identities within a given style. In order to derive 3D cues from stylized images with sparse geometric information, we adopt a progressive training strategy consisting of three stages: geometry warm-up for mesh deformation, joint fine-tuning of shape and texture, and further texture refinement. To stabilize face diversity of the original 3DMM during stylization, we introduce a Consistent Displacement Loss (CDL), which promotes coherent mesh deformation across diverse identities and facilitates the generalization of stylization under diverse shape conditions.

Our contributions can be summarized as follows:
\begin{itemize}
    \item We formalized the requirements for a stylized 3DMM and introduced the first method to fulfill them by fine-tuning the realistic face generator in a three‑stage process.
    \item We developed Explicit Attribute-preserving Stylization (EAS), a new facial image stylization pipeline using adjusted noise initialization.
    \item We proposed the Explicit Attribute-preserving Module (EAM), which integrates facial landmarks, rotation, and expression into an existing diffusion model to preserve desired attributes during generation.
    \item We formulated a novel CDL loss that enhances identity-level face diversity by preventing mode collapse.
\end{itemize}

\section{Related Work}

\subsection{Morphable Face Model}
Morphable face models, introduced by Blanz and Vetter~\cite{blanz1999morphable}, represent faces using dense geometry and texture modeled via principal component analysis (PCA), enabling intuitive control over attributes such as gender and fullness. 
The Basel Face Model~\cite{paysan20093d}—subsequently expanded to a dataset of 10,000 facial scans~\cite{booth20163d,booth2018large}—laid the foundation for modern face modeling. Subsequently,  FaceWarehouse~\cite{cao2013facewarehouse} introduced multi-linear models to capture identity and expression variations. More recently, FLAME~\cite{li2017learning} and ICT-FaceKit~\cite{li2020learning} have leveraged pose-dependent corrective blendshapes and non-linear expression models, derived from larger facial-scan datasets, to further refine facial expressions. In this work, we utilize FLAME as the source model, distill it into a learning-based model, and then fine-tune it to obtain a new morphable model.

\subsection{Text to 3D Face Generation}
Recent advancements in text-to-3D generation, driven by progress in text-to-image models~\cite{rombach2022high,saharia2022photorealistic} and vision-language models~\cite{radford2021learning}, have demonstrated remarkable capabilities. One line of research leverages neural fields such as NeRF~\cite{mildenhall2021nerf}, 3DGS~\cite{kerbl20233d}, or DMTet~\cite{shen2021deep}. Despite their success in producing high-quality stylized characters~\cite{abdal20233davatargan,zhang2023styleavatar3d,huang2024humannorm,liu2024headartist,zhou2024headstudio} via multi-view rendering, generating meshes with a consistent structure remains challenging, limiting compatibility with existing graphics pipelines.

Another line of work builds upon established face models such as FLAME and ICT-FaceKit, achieving high-quality textures while preserving compatibility with standard graphics pipelines, owing to their consistent mesh structure. For instance, CLIPface~\cite{aneja2022clipface} generates texture maps for a given FLAME model, while DreamFace~\cite{zhang2023Dreamface} and UltraAvatar~\cite{zhou2024ultravatar} sample or estimate blendshape parameters to obtain geometry while generating the textures. These methods confine shapes within the realistic 3DMM space, restricting geometric exaggerations or abstractions beyond the distribution of scanned-face datasets.

\subsection{Surface Deformation Network}
Learning implicit functions for 3D shapes has proven highly effective in representing complex geometries~\cite{park2019deepsdf,mescheder2019occupancy,michalkiewicz2019implicit,lipman2021phase}. DIF-Net~\cite{deng2021deformed} uses MLPs to learn a signed distance function (SDF) alongside a volumetric deformation function, while DD3C~\cite{jung2022deep} and LeGO~\cite{yoon2024lego} adopt surface deformation for face stylization. Although DD3C and LeGO demonstrate the creation of impressive stylization results, both rely on ground-truth target meshes created by skilled artists for training, thereby limiting accessibility for novice users. Building on their finding that surface deformation is well-suited for facial modeling, we advance this approach by eliminating the need for ground-truth target meshes. Specifically, we use text descriptions to generate target images and then deform the surface mesh accordingly.

\subsection{Text-Based Portrait Stylization}
Text-based portrait stylization focuses on stylizing a given image while preserving the subject’s identity. Both GAN-based~\cite{gal2022stylegan,patashnik2021styleclip} and diffusion-based~\cite{li2024photomaker,ye2023ip,wang2024instantid} methods have made notable strides by leveraging pre-trained generative models to balance style and identity. Unfortunately, these approaches are not designed for 3D deformation, making it difficult to guarantee precise alignment with the source image. Therefore, we propose an explicitly aligned stylization approach that ensures not only identity and style preservation but also spatial alignment and retention of the subject’s expressions.



\section{Methods}

\begin{figure}[h]
    \hspace{-3mm}
    \includegraphics[width=1.04\linewidth]{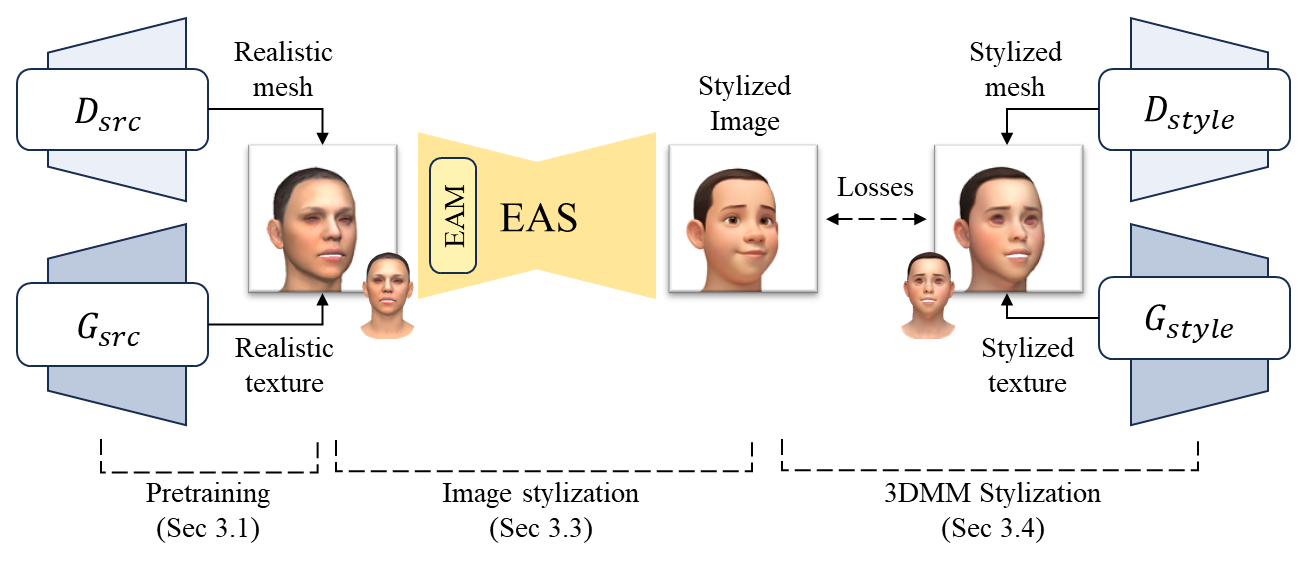} \vspace{-2mm}
    \caption{StyleMM fine-tunes Deformation network $D_{src}$ and texture generator $G_{src}$ pre-trained on FLAME into stylized models $D_{style}$ and $G_{style}$, respectively using Explicit Attribute-preserving Stylization (EAS). In this process, the Explicit Attribute-preserving Module (EAM) is a component of EAS that enables the preservation of alignments.}\vspace{-2mm}    \label{fig:overview}\vspace{-3mm}    
\end{figure}

Our goal is to build a 3DMM for stylized faces, which reflects the style described by a user-provided text prompt.
To achieve this, we fine-tune two pre-trained networks $D_{src}$ and $G_{src}$, both originally trained on natural human 3DMM faces (Section~\ref{sec:pretrained_networks}). 
According to the given style description, our framework automatically generates a dataset of stylized facial images (Sections~\ref{sec:eam} and~\ref{sec:stylization}) and fine-tunes the networks by comparing rendered results with the image data (Section~\ref{sec:finetuning}). After training, the resulting models, denoted as $D_{style}$ and $G_{style}$, operate together as a stylized 3DMM conditioned on shape and texture, respectively. An overview of the StyleMM training process is shown in Figure~\ref{fig:overview}.

\subsection{Pre-trained Networks}\label{sec:pretrained_networks}

\begin{figure*}[t]
    \centering
    \vspace{-2mm}
    \includegraphics[width=\linewidth]{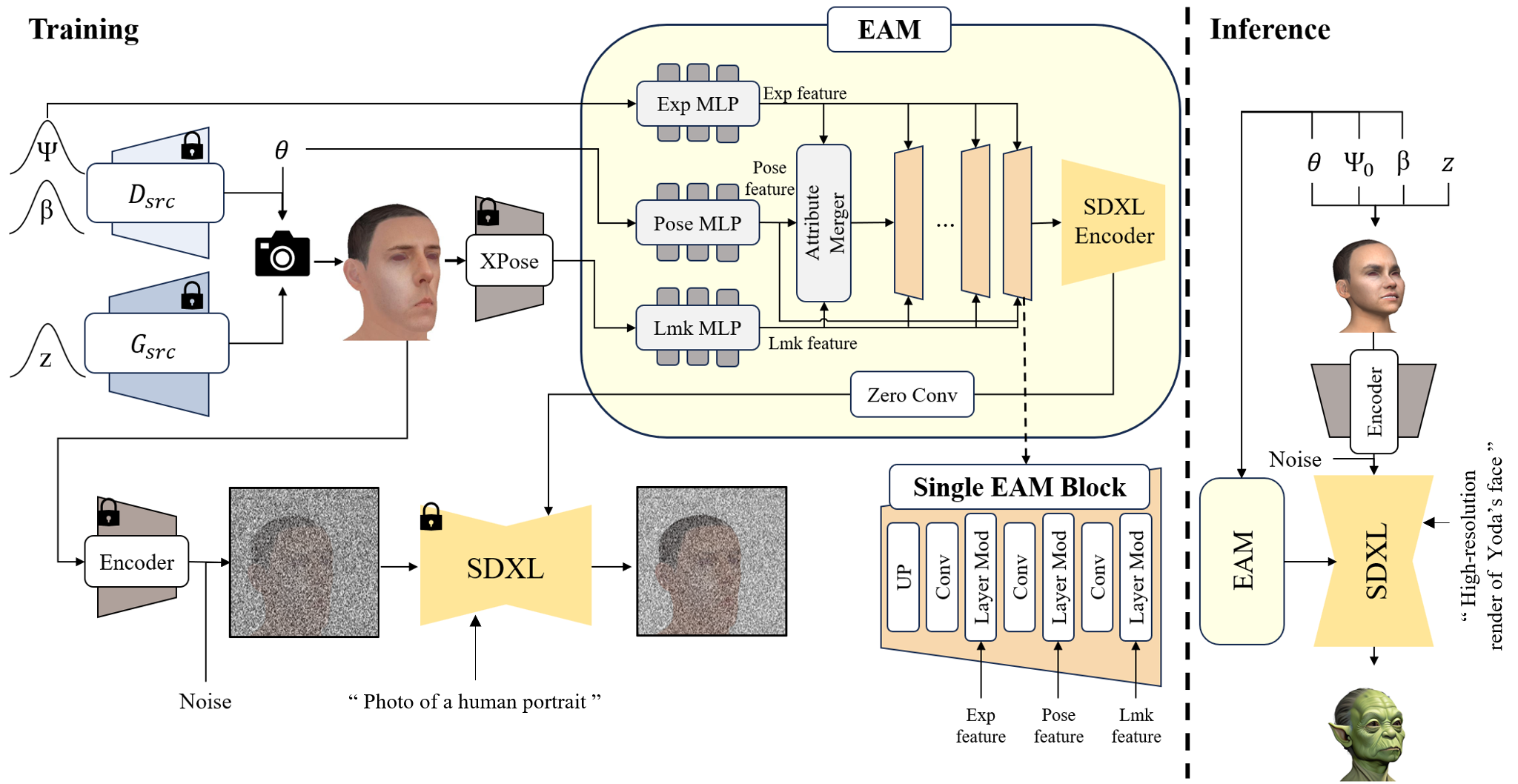} \vspace{-5mm}
    \caption{\textbf{Overview of Explicit Attribute-preserving Stylization.} Left: training process of EAM. Right: Inference process of EAS, equipped with EAM.}\label{fig:EAS} \vspace{-3mm}
\end{figure*}

\subsubsection{Surface Deformation Network}
Recent studies have shown that a surface deformation network can efficiently modify geometry while preserving the underlying mesh structure in face stylization~\cite{jung2022deep,yoon2024lego}. Therefore, we adopt the surface deformation network from LeGO~\cite{yoon2024lego}, which deforms a template face mesh using FLAME shape $\boldsymbol{\beta}$ and expression $\boldsymbol{\psi}$ parameters. Specifically, given vertices on a template mesh $\boldsymbol{V}_0 \in \mathbb{R}^{N\times 3}$ ($N$ is the number of vertices), the network outputs per-vertex offsets $\Delta \boldsymbol{V}$ such that the final mesh is $\boldsymbol{V} = \boldsymbol{V}_0 + \Delta \boldsymbol{V}$.

This pre-trained model $D_{src}$ provides the geometry of a 3DMM capable of generating diverse face shapes and expressions while maintaining consistent vertex connectivity.

\subsubsection{Texture Generator}
For texture synthesis, we utilize a pre-trained StyleGAN2~\cite{karras2020analyzing} model capable of generating textures in UV space. Specifically, a latent code $\boldsymbol{z}$ is first mapped to an intermediate style vector $\boldsymbol{w}$, which then modulates the convolution layers to generate a high-resolution RGB texture. Originally trained on natural human face datasets~\cite{bai2023ffhq,karras2019style}, the model, denoted as $G_{src}$, can synthesize detailed facial textures.

\subsection{Explicit Attribute-preserving Module}\label{sec:eam}
For image-based fine-tuning, our framework constructs a stylized image dataset by pairing each stylized image with its corresponding source image based on user-defined style text. Specifically, we render natural human faces using pre-trained networks $D_{src}$ and $G_{src}$, and apply the desired style using EAS. Although previous portrait stylization methods are well-suited for both stylization and identity preservation~\cite{ye2023ip,li2024photomaker}, they often introduced changes in alignment or expression in our experiments (See Figure~\ref{fig:i2i_comparison}). Such inconsistencies can hinder building a 3D face model due to unaligned deformation targets.

To address this, we propose EAS, which is composed of SDXL~\cite{podell2023sdxl} and a newly proposed Explicit Attribute-preserving Module (EAM) as shown in Figure~\ref{fig:EAS}. The EAM provides explicit conditions to SDXL to preserve attributes of the source face, such as scale, rotation, alignment, and expression. The EAM consists of three condition-encoding MLPs, an Attribute Merger, a series of EAM blocks, and an SDXL Encoder. The weights of the SDXL Encoder are initialized using those of the pre-trained SDXL, following the approach of ControlNet~\cite{zhang2023adding}. Sparse landmarks ($lmk$), head rotation ($\theta$), and facial expression ($\psi$) are passed through their respective encoding MLPs, merged before being fed to a series of EAM Blocks. Each EAM block accepts the processed feature as input, upsample them, and feed them to convolutional layers equipped with Adaptive Layer Normalization initialized with zero~\cite{peebles2023scalable}.

To train the EAM, we randomly sample shape parameters $\boldsymbol{\beta}$, expression parameters $\boldsymbol{\psi}$, and texture latent codes $\boldsymbol{z}$ from normal distribution, which represent a source human face. We then render the output faces with random rotation angles $\theta$. Both $\theta$ and $\boldsymbol{\psi}$ are directly used as inputs to the EAM during training, while $lmk$ is extracted using a pre-trained network~\cite{yang2024x}. We use a sparse subset of five landmarks—two from the eyes, one from the nose, and two from the lips—to guide alignment between the stylized face and the source image. This selection allows stylization of facial features while avoiding constraints on outer facial shape (e.g., the chin), which is often meant to be stylized. The training loss for the EAM can be expressed as follows:
\begin{equation}
\mathcal{L}_{\text{train}} = \mathbb{E}_{x_0, t, txt,lmk, \theta, \psi, \epsilon \sim \mathcal{N}(0,1)} \left[ \| \epsilon_\theta\left(x_t, t, txt, lmk, \theta, \psi\right) - \epsilon \| _2^2 \right],
\end{equation}

where $\epsilon$ is the noise added during diffusion training, and $t$ is the denoising timestep. $x_t$ is the noise-added image at timestep $t$ and $txt$ is the text input. Here, we train the EAM with full and partial conditions. For the full condition, we use all three inputs-$lmk, \theta, \Psi$-while for the partial condition, only two input parameters are used. We omit each conditional parameter in 25\% of the training samples. This hybrid training scheme enables the EAM to learn without relying solely on any one parameter, while still effectively leveraging all three when available.



\subsection{Explicit Attribute-preserving Stylization}\label{sec:stylization}

We use the trained EAM during inference to stylize rendered faces while preserving their translation, rotation, and expression. To further accelerate inference and preserve the original structure and identity, we initialize the latent variable using noise added source image $x_t$ instead of generating images from random noise. Here, $t$ is set to 19 out of total 25 DDIM~\cite{song2020denoising} sampling. This adjusted initialization speeds up inference without degrading stylization quality. This is because the early denoising steps focus primarily on coarse geometric structure, while style and details emerge during the middle to later denoising steps~\cite{meng2021sdedit,kim2023datid}. We observed that setting this $t$ lower (e.g., $ t =10$) led to faster inference speed but significantly decreased stylization capability as shown in Figure~\ref{fig:init_ablation}. The resulting stylized images, paired with the corresponding rendered source images, serve as the synthetic paired data for our image-based fine-tuning process.

\subsection{Image-Based Learning for 3DMM Stylization} \label{sec:finetuning}

\begin{figure*}[ht!]
    \centering
    \includegraphics[width=1\linewidth]{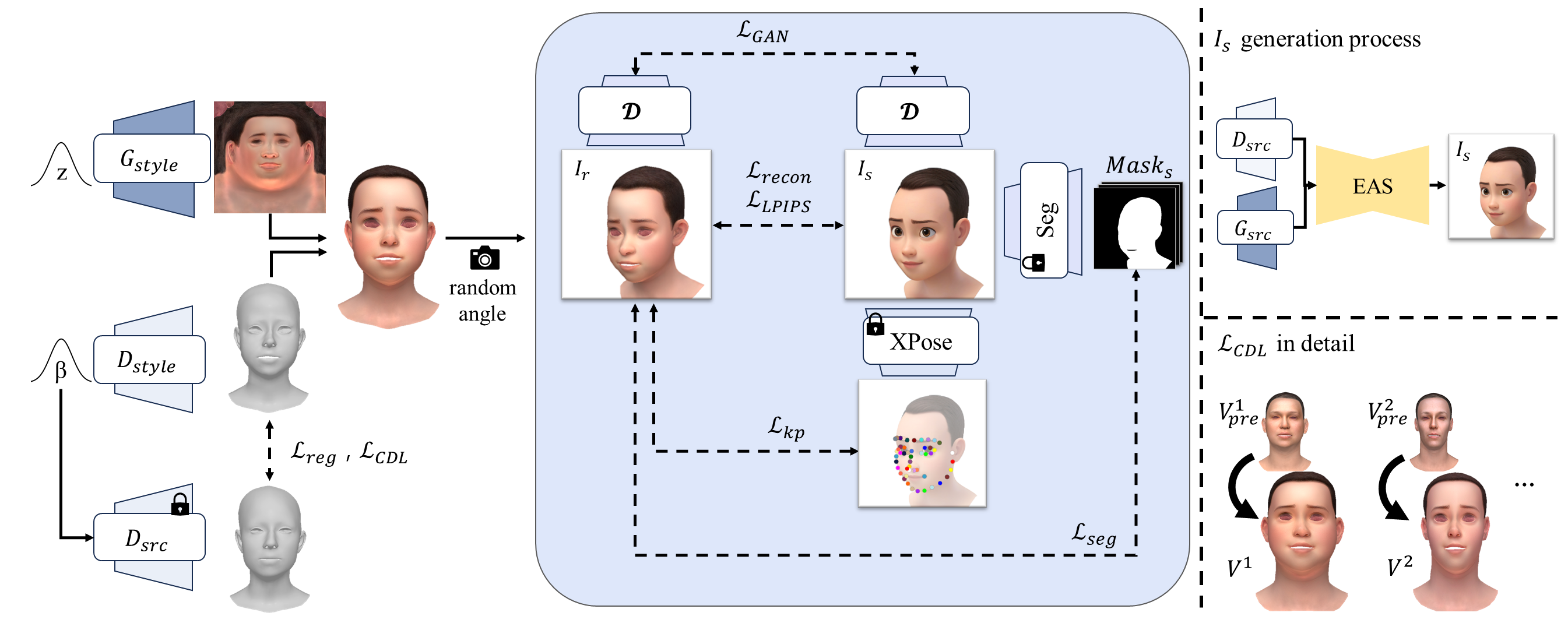}
\caption{\textbf{Overview of the StyleMM training pipeline.}
Our method stylizes a 3D morphable face model through a three-stage process that leverages distinct loss functions. A deformation network $D_{\text{style}}$ and a texture generator $G_{\text{style}}$ are optimized using style-supervised 2D image pairs rendered using random shapes, textures, and viewpoints. The proposed Consistent Displacement Loss $\mathcal{L}_{\text{CDL}}$ encourages locally consistent deformation patterns across different identities as shown in the bottom right of the figure.}\label{fig:enter-label} \vspace{-3mm}
\end{figure*}

To align the rendered output from the 3D network with the stylized facial images generated by EAS, which are generated under the same conditions, we fine-tune both the surface deformation network and the texture generator. In this process, we found that a single reconstruction loss is insufficient because the 2D supervision provides limited geometric information. Therefore, we divide the training into a three-stage process: an initial geometry warm-up phase, joint fine-tuning of shape and texture, and the final stage for texture refinement. An overview of this training pipeline is illustrated in Figure~\ref{fig:enter-label}. We will introduce loss terms for the stylization and deformation stabilization in subsections ~\ref{sec:adaptation} and ~\ref{sec:stabilization} respectively.

\subsubsection{Style Adaptation Losses} \label{sec:adaptation}
To guide the model toward the target, we apply different loss terms at various training stages, each extracting meaningful cues from 2D images and transferring them to the 3D face model. 

\paragraph{Geometry Warm-Up.}
Recall that $D_{src}$ and $D_{style}$ are the surface deformation networks that take the canonical 3D face geometry as input and predict per-vertex offsets to produce the human face and the stylized face, respectively. In this warm-up stage, we finetune $D_{src}$ to become $D_{style}$ to establish an accurate geometric foundation before jointly training $D_{style}$ and $G_{style}$. Instead of relying on raw pixel values, which can be heavily influenced by textures, we employ 2D keypoint matching as a more stable guide for initial shape alignment. Specifically, we extract a set of facial landmarks from each stylized image using an off-the-shelf detector X-Pose ~\cite{yang2024x}, and assign the corresponding 3D vertices in the mesh. By projecting these vertices onto the screen space, the discrepancies between the projected vertices and the detected 2D landmarks are penalized. 
\begin{equation}
\begin{aligned}
\mathcal{L}_{\mathrm{kp}} = &\sum_{i} \Bigl\|\Pi(\boldsymbol{v}_i)-\boldsymbol{k}_i\Bigr\| 
\end{aligned}
\end{equation}
where \(\Pi(\cdot)\) denotes the camera projection function, \(\boldsymbol{v}_i \in \mathbb{R}^3\) are the 3D vertices, \(\boldsymbol{k}_i \in \mathbb{R}^2\) are the detected 2D landmarks. This approach encourages $D_{style}$ to capture a coarse geometric structure, mitigating potential distractions from complex stylized textures.

\paragraph{Joint Fine-Tuning.}
After the warm-up, we jointly fine-tune the surface deformation network $D_{style}$ and the texture generator $G_{style}$ on the style data from EAS. The joint training combines geometric and textural information, ensuring improved reconstruction fidelity and style adherence.
We employ a reconstruction loss between the rendered image $\boldsymbol{I}_r$ and the stylized image $\boldsymbol{I}_s$:
\begin{equation} \label{eq:recon}
\mathcal{L}_{\text{recon}} = \| \boldsymbol{I}_r - \boldsymbol{I}_s \|_2 + \lambda_{CLIP} \cdot CLIP{(\boldsymbol{I}_r, \boldsymbol{I}_s)}+\lambda_{DINO} \cdot DINO{(\boldsymbol{I}_r, \boldsymbol{I}_s)}
\end{equation}
where $CLIP$ denotes the cosine similarity in CLIP~\cite{radford2021learning} embedding space, $DINO$ denotes the cosine similarity of DINOv2~\cite{oquab2023dinov2} feature, and $\lambda_{CLIP}$ and $\lambda_{DINO}$ represent the weighting factor for each reconstruction loss, respectively. The reconstruction loss encourages the rendered image to closely match the stylized target in terms of geometry and texture.

Because the structural and textural information are entangled in the image domain, the reconstruction loss alone often fails to provide accurate geometric detail, especially for fine-grained facial parts. To address this, we employ a segmentation-guided alignment loss computed over facial part segmentation maps, which explicitly guides the model to match the spatial layout of key regions.
\begin{equation}
\mathcal{L}_{\text{seg}} = \sum_{c \in \mathcal{C}} \| \boldsymbol{Mask}^{(c)}_r - \boldsymbol{Mask}^{(c)}_s \|_2.
\quad
\end{equation}
Here, $\boldsymbol{Mask}^{(c)}_r$ and $\boldsymbol{Mask}^{(c)}_s$ denote the segmentation masks of class $c$ (eye, nose, ear, and background) from the rendered and stylized images, respectively. Note that this segmentation is generated using
our stylized-face segmentation network which will be explained in Section~\ref{sec:segmentation}.



\paragraph{Texture Refinement.}

In the final stage, we enhance the texture generator to improve the quality of reproduced fine-grained textural details from the style images. We employ a perceptual similarity loss based on LPIPS~\cite{zhang2018unreasonable} for local fidelity.
\begin{equation}
\mathcal{L}_{\text{LPIPS}} = \| \phi(\boldsymbol{I}_r) - \phi(\boldsymbol{I}_s) \|_2,
\end{equation}
where $\phi(\cdot)$ denotes a LPIPS feature extractor.

To further enhance global texture plausibility, we incorporate an adversarial loss with an additional trained discriminator.
\begin{equation}
\mathcal{L}_{\text{GAN}} = \log \mathcal{D}(\boldsymbol{I}_s) + \log (1 - \mathcal{D}(\boldsymbol{I}_r)),
\end{equation}
where $\mathcal{D}(\cdot)$ denotes the discriminator. This adversarial loss encourages the generator to globally align the generated face with target style distribution when rendered.

\subsubsection{Deformation Stabilization Losses} \label{sec:stabilization}
Because the style adaptation losses are primarily guided by 2D observations, they can produce unstable or collapsed deformations when applied to 3D models. To address this, we introduce complementary loss terms that stabilize the deformation process and preserve plausible facial geometry throughout training.

Our goal is to generalize the stylization across the entire shape parameter space, while preserving the structural diversity of the original 3DMM.
{In our experiments, finetuning the 3D networks with stylized images—each depicting a different individuals and view—often resulted in convergence to a single dominant geometric form, limiting identity diversity. This may be caused by the inherent randomness of the diffusion model, such that regardless of facial attribute alignment, intensity or visual interpretation of the style may vary at each inference.} To address this, we propose a Consistent Displacement Loss, denoted as $L_{CDL}$. $L_{CDL}$ encourages coherent deformation by aggregating partial cues into consistent displacement patterns across identities in a batch.
\begin{equation}
    \label{eq:def_reg}
    \mathcal{L}_{\text{CDL}} =
    \mathrm{Var} \left( \left\{ \mathrm{vertex\_tangent} \left( \boldsymbol{V}^{(i)} - \boldsymbol{V}_{\text{pre}}^{(i)} \right) \right\}_{i=1}^{B} \right)
\end{equation}
where $\mathrm{vertex\_tangent}(\cdot)$ denotes projection onto the local tangent space of each vertex, and $\mathrm{Var}(\cdot)$ computes the variance across a batch of $B$ samples with different shape parameters. By aligning style deformation in the parameter space, $\mathcal{L}_{\text{CDL}}$ helps preserve the shape diversity of the original 3DMM.

In addition, we apply an auxiliary regularization loss to preserve the pretrained network’s knowledge of plausible face geometry, including vertex positions, surface normals, and internal face angles.
\begin{equation} \label{eq:regeq}
\begin{aligned}
\mathcal{L}_{\mathrm{reg}} =
&\ \lambda_v\,
\bigl\|
\boldsymbol{V} - \boldsymbol{V}_{\mathrm{pre}}
\bigr\|_2^2
+ \lambda_n\,
\bigl\|
\boldsymbol{N} - \boldsymbol{N}_{\mathrm{pre}}
\bigr\|_2^2 \\
&\ + \lambda_{\mathrm{ang}} \sum_{t\in\mathcal{T}} \sum_{\alpha \in \{\alpha_t^1, \alpha_t^2, \alpha_t^3\}}
\bigl( \cos(\alpha) - \cos(\alpha^{\text{pre}}) \bigr)^2
\end{aligned}
\end{equation}
where $\boldsymbol{V}$, $\boldsymbol{N}$, and $\boldsymbol{\alpha}$ denote the current vertex positions, normals, and internal face angles produced by $D_{style}$ given a shape parameter $\boldsymbol{\beta}$ and expression parameter $\boldsymbol{\psi}$. $\boldsymbol{V}_{\mathrm{pre}}$, $\boldsymbol{N}_{\mathrm{pre}}$, and $\boldsymbol{\alpha}_{\mathrm{pre}}$ denote the corresponding outputs from the pre-trained model $D_{src}$ under the same parameters.

\subsubsection{Overall Training Objective}

Based on the style adaptation and deformation stabilization losses introduced in the previous sections, we define the overall training objective for each stage of the stylization process. Here, each $\lambda$ value is a weighting factor.

\textbf{Geometry Warm-up:}
\begin{equation}\label{eq:geometry-warmup}
\mathcal{L}_{\mathrm{warm}} =
\lambda_{\mathrm{kp}} \, \mathcal{L}_{\mathrm{kp}} +
\lambda_{\mathrm{reg}} \, \mathcal{L}_{\mathrm{reg}} +
\lambda_{\mathrm{CDL}} \, \mathcal{L}_{\mathrm{CDL}}.
\end{equation}

\textbf{Joint Fine-tuning:}
\begin{equation}\label{eq:joint-finetuning}
\mathcal{L}_{\mathrm{joint}} =
\lambda_{\text{recon}} \, \mathcal{L}_{\text{recon}} +
\lambda_{\text{seg}} \, \mathcal{L}_{\text{seg}} +
\lambda_{\mathrm{reg}} \, \mathcal{L}_{\mathrm{reg}} +
\lambda_{\mathrm{CDL}} \, \mathcal{L}_{\mathrm{CDL}}.
\end{equation}

\textbf{Texture Refinement:}
\begin{equation}\label{eq:texture-refinement}
\mathcal{L}_{\mathrm{tex}} =
\lambda_{\text{LPIPS}} \, \mathcal{L}_{\text{LPIPS}} +
\lambda_{\text{GAN}} \, \mathcal{L}_{\text{GAN}}.
\end{equation}
The style adaptation losses are applied differently at each of the three steps. Specifically, $\mathcal{L}_{kp}$ is applied during the geometry warm-up stage, $\mathcal{L}_{recon}$ and $\mathcal{L}_{seg}$ are applied during the joint fine-tuning stage, and $\mathcal{L}_{LPIPS}$ and $\mathcal{L}_{GAN}$ are applied during the texture refinement stage. In contrast, both of the deformation stabilization losses are employed  during the warm-up and joint fine-tuning stages.

\subsubsection{Image Segmentation}\label{sec:segmentation}
Facial part segmentations are required to apply the segmentation-guided alignment loss on stylized images. The segmentations are also used for image masking to exclude regions such as the inner mouth, eyes, and background, which are irrelevant to stylized 3DMM construction. Because no existing method can accurately mask the desired parts of diverse stylized faces, and general segmentation models~\cite{kirillov2023segment} cannot segment detailed areas, we introduce a simple feature-based masking network $f_{\text{mask}}$ for stylized faces, trained in a few-shot manner. We manually create five facial images for each of five different styles, resulting in 25 training pairs. Each pair is annotated with six segments: eyes, nose, ears, mouth, face, and outer face including hair and background.

For efficient training, we use features from pre-trained models~\cite{xie2023high, oquab2023dinov2, darcet2023vision}, known for their efficacy in downstream tasks~\cite{zhang2023tale,yun2024stylized, tang2023emergent, yun2024representative}, by fusing and decoding them into a mask using a zero-initialized Residual Convolutional network~\cite{he2016identity}. Specifically, we fuse the features from Stable Diffusion~\cite{rombach2022high} and DINOv2~\cite{oquab2023dinov2}, inspired by SD-DINO\cite{zhang2023tale}. The training is conducted with the cross-entropy loss $\mathcal{L} = -\frac{1}{N_{mask}} \sum_{i=1}^{N_{mask}} \log \Bigl( \hat{y}_{i}, y_i \Bigr)$, where $\hat{y}$ is the estimated mask label, $y$ is the ground-truth mask label, and $N_{mask}$ is a total number of mask labels. After training $f_{\text{mask}}$, we obtain a mask $\boldsymbol{M}$ for each stylized image and use them during joint fine-tuning and texture refinement process. For a mesh, we define the part regions in the UV space, render masks separately, and apply them to ensure consistency.

\section{Experiments}

\subsection{Implementation Details}\label{sec:implementation}
{Training EAM and EAS inference} were conducted on a system with a single RTX A6000 GPU while StyleMM were trained and evaluated on a system with a single RTX 3090 GPU. The EAM was trained for a total of 20,000 iterations with a learning rate of 1e-5. StyleMM was trained on 10,000 paired images with a learning rate of 5e-5 for both the surface deformation network and the texture generator. Warmup training was performed for 2 epochs with the weighting factors $\lambda_{kp}$, $\lambda_{reg}$, and $\lambda_{CDL}$ from Eq.~\eqref{eq:geometry-warmup} set to 300, 4, and 4000, respectively. Joint fine-tuning was performed for 4 epochs with the weighting factors $\lambda_{recon}$, $\lambda_{seg}$, $\lambda_{reg}$, and $\lambda_{CDL}$ from Eq.~\eqref{eq:joint-finetuning} set to 500, 100, 2, and 500 respectively. Texture refinement was conducted for 4 epochs with the weighting factors $\lambda_{LPIPS}$ and $\lambda_{GAN}$ from Eq.~\eqref{eq:texture-refinement} set to 50 and 1, respectively. For the sub-losses in Eq.\eqref{eq:recon}, $\lambda_{\text{CLIP}}$ was fixed at 0.2, while $\lambda_{\text{DINO}}$ was set to 0.2 initially and disabled after 500 iterations. The regularization weights $\lambda_v$, $\lambda_n$, and $\lambda_{ang}$ in Eq.~\eqref{eq:regeq} were set to 1, 50, and 1000, respectively. The 3D pipeline required about 3 hours for training.

\subsection{Qualitative Results}\label{sec:qualitative}

\begin{figure*}
    \centering
    \includegraphics[width=0.9\linewidth]{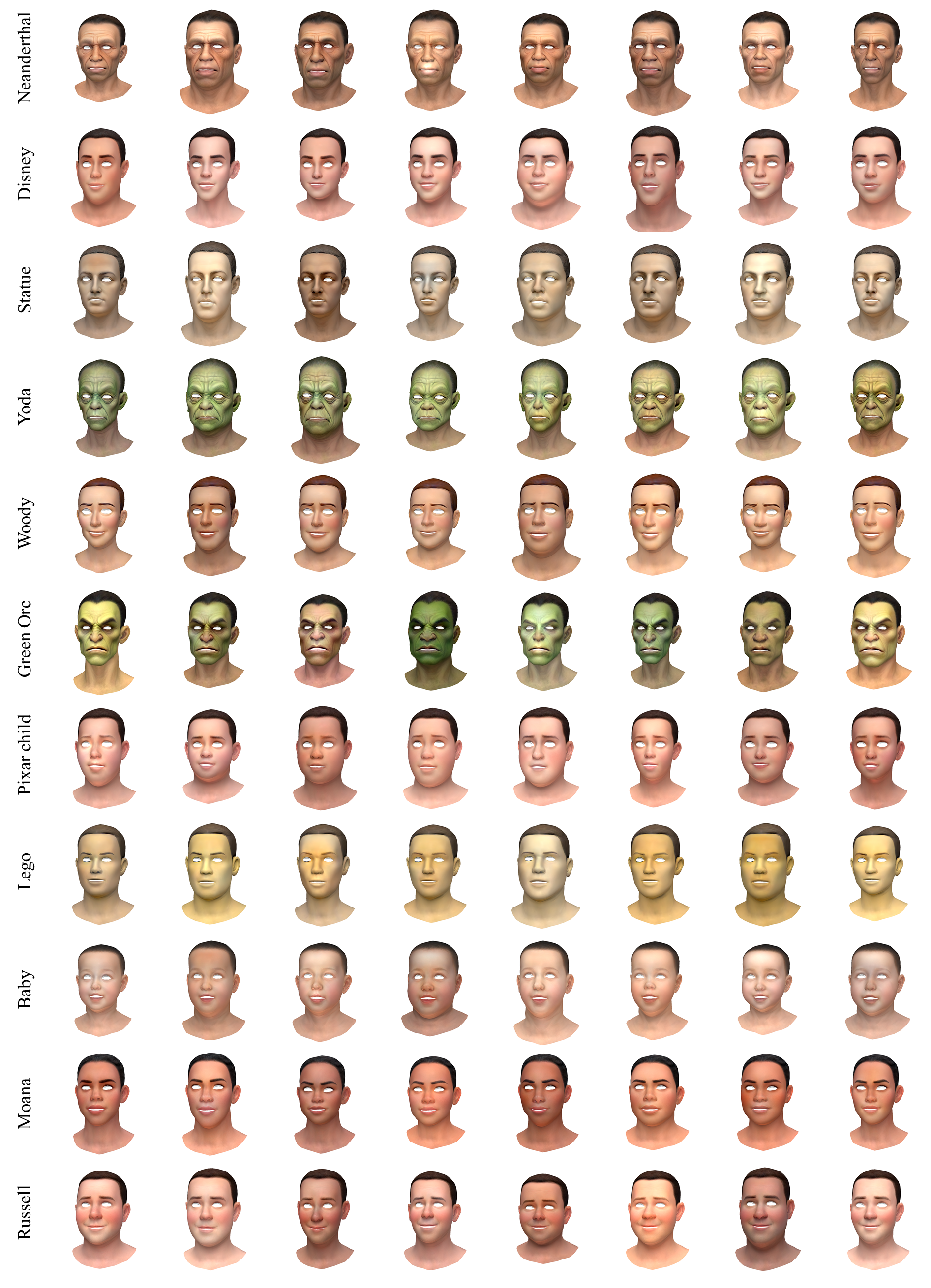}
    \vspace{-2mm}
    \caption{Results of StyleMM, shape and textures are randomly sampled from $D_{style}$ and $G_{style}$.}
    \label{fig:ours_all}
\end{figure*}

We evaluated our approach across three fundamental elements of a stylized 3DMM: (1) maintained correspondence, (2) disentangled control, and (3) stylization beyond realistic geometry and texture. Additional results are presented in Figure~\ref{fig:ours_all}.

\subsubsection{Maintained Correspondence}

\begin{figure}[t]
    \centering
    \vspace{-2mm}
    \includegraphics[width=0.95\linewidth]{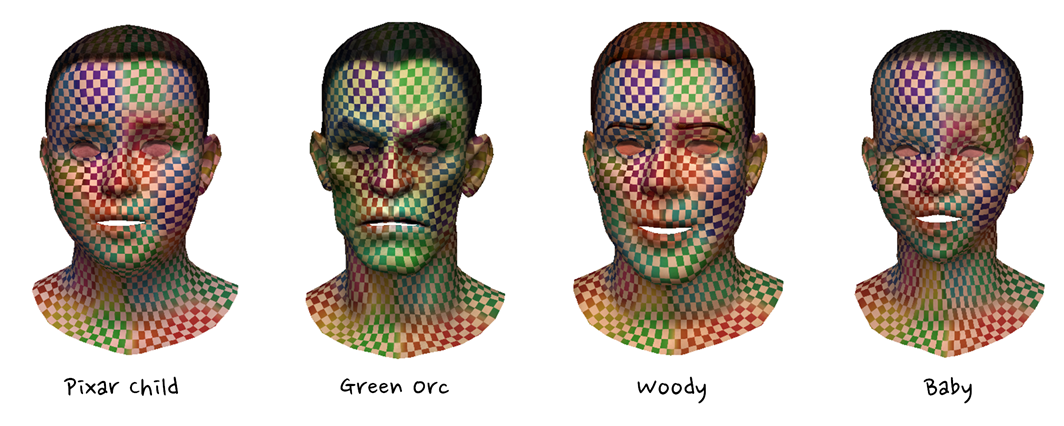}
    \vspace{-4mm}
    \caption{Stylized faces rendered with the same UV pattern to visualize vertex correspondence. Key regions remain aligned across identities.}\vspace{-3mm}
    \label{fig:correspondance}
\end{figure}

We first assessed whether the generated faces preserve dense point-to-point correspondences across different identities and styles. To visualize vertex alignment, we overlaid the same UV texture to several stylized outputs. As illustrated in Figure~\ref{fig:correspondance}, key facial features consistently occupy the same semantic regions across different stylized faces. 
This demonstrates that our framework ensures reliable vertex indexing, which is essential for downstream tasks such as animation retargeting, texture editing, and semantic part manipulation.
 
\subsubsection{Disentangled Control}

\begin{figure*}[t]
    \centering \vspace{-3mm}
    \includegraphics[width=0.9\linewidth]{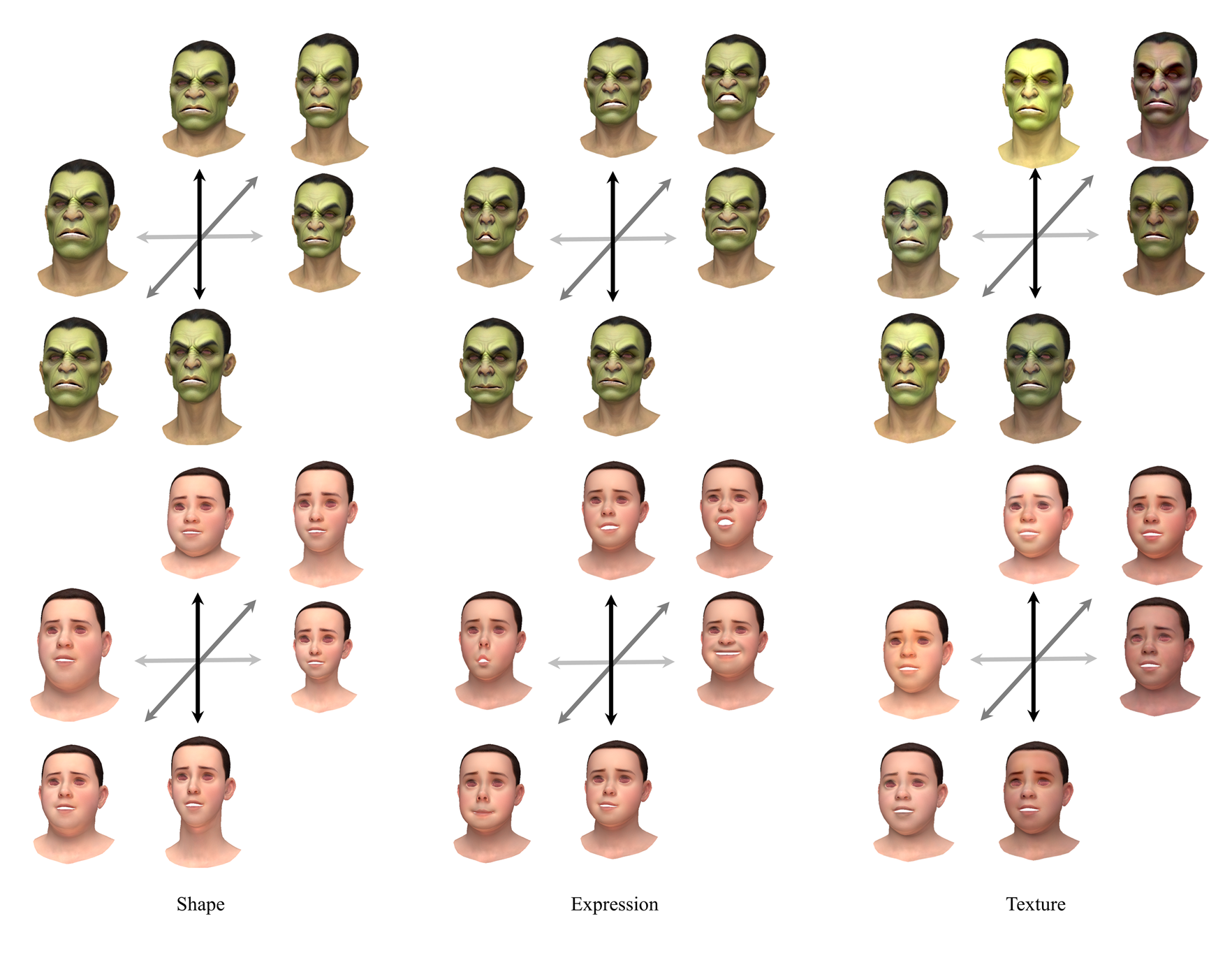}
    \vspace{-6.5mm}
    \caption{Disentangled parametric control of shape (left), expression (middle), and texture (right), with other factors fixed.}\vspace{-4mm}
    \label{fig:control}
\end{figure*}

To demonstrate that our stylization model functions as a 3DMM, we present parametric control results for shape, expression, and texture. Figure~\ref{fig:control} visualizes this through the three sets of variations. The left block illustrates shape variations. It shows that the model inherits a structured control space from the pretrained 3DMM, where variations align with prominent facial features. The middle block demonstrates expression-level control. Because our framework maintains vertex-level correspondence even with pretrained 3DMM faces, it supports the direct reuse of expression blendshape basis defined in the original 3DMM. 
The right block shows diverse texture variations. The generated textures maintain identity-relevant details inherited from the latent space of the pretrained texture generator, while allowing for diverse stylization. Overall, StyleMM enables independent parametric control over key facial attributes, consistent with the original 3DMM.

\begin{figure}
    \centering
    \includegraphics[width=0.95\linewidth]{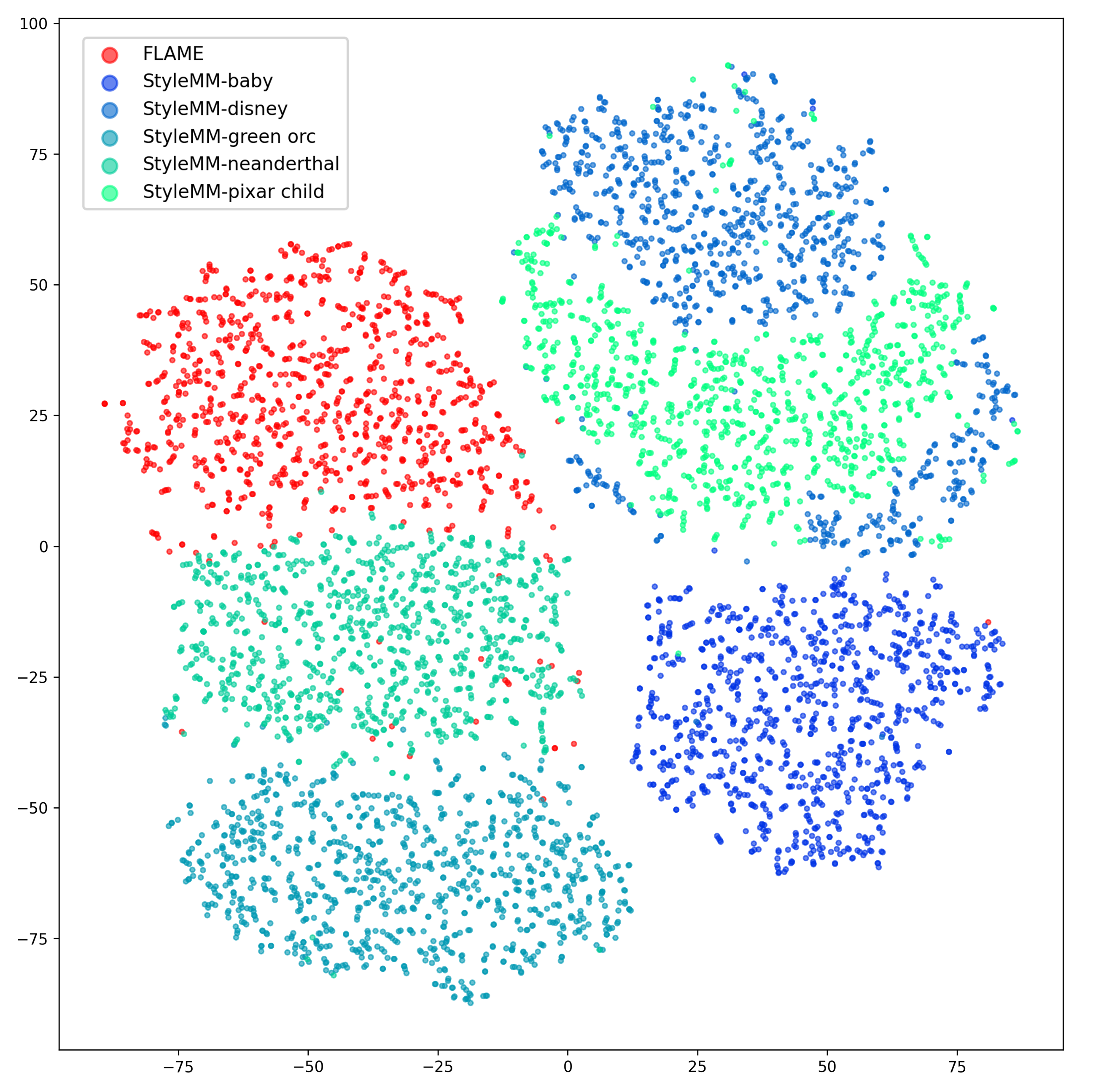}
    \caption{The t-SNE projection of randomly sampled faces for each style. Our model extends beyond the realistic representation, distinctly diverging from the representation space of FLAME.}\vspace{-2mm}
    \label{fig:beyond}
\end{figure}

\begin{figure}[t]
    \centering
    \includegraphics[width=\linewidth]{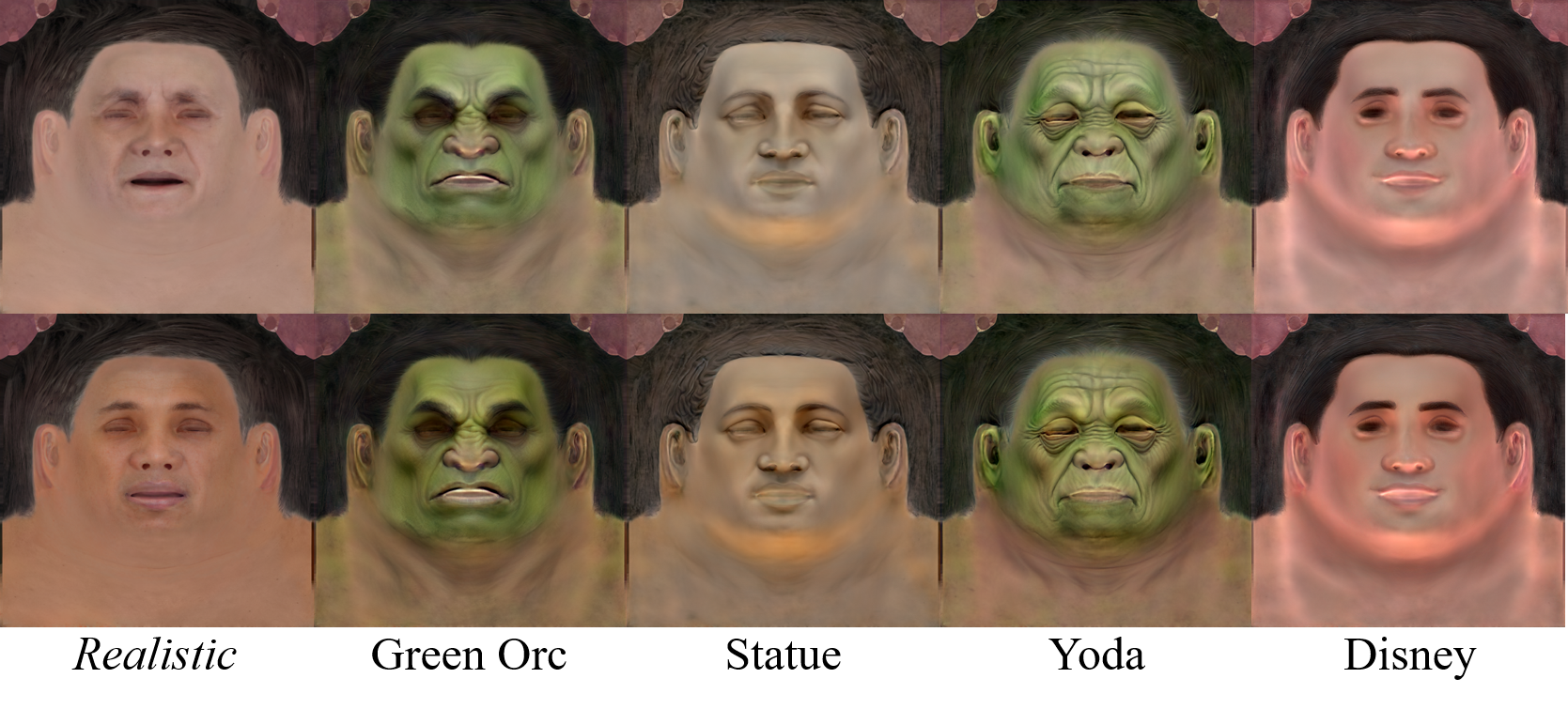}\vspace{-2mm}
    \caption{Generated Texture from $G_{src}$ (Realistic) and variants of $G_{style}$. Each row shares the same texture code.}\vspace{-6mm}
    \label{fig:beyond_texture}
\end{figure}

\subsubsection{Stylization Beyond Realistic Geometry and Texture}
To examine the expressive capacity of our stylized 3DMM beyond the realistic 3DMM, we visualize the distribution of generated identities based on their mesh geometry.  
We randomly sampled a total of 1,000 face meshes across different styles and applied t-SNE to their vertex coordinates to project them into a 2D space.
As shown in Figure~\ref{fig:beyond}, stylized identities form distinct clusters that extend beyond the region occupied by the original FLAME model.  
This suggests that our framework can generate geometries that surpass the anthropomorphic bounds of conventional 3DMMs, enabling more expressive and creative stylization. We also demonstrate expressive capacity in texture generation by visualizing the generated results in Figure~\ref{fig:beyond_texture}. While each row shares the same texture code, the results show that $G_{style}$ produces textures with expressiveness that surpasses realistic appearances.

\subsection{Comparison with Baselines}\label{sec:qualitative}
\begin{figure*}
    \centering \vspace{-2mm}
    \includegraphics[width=0.97\linewidth]{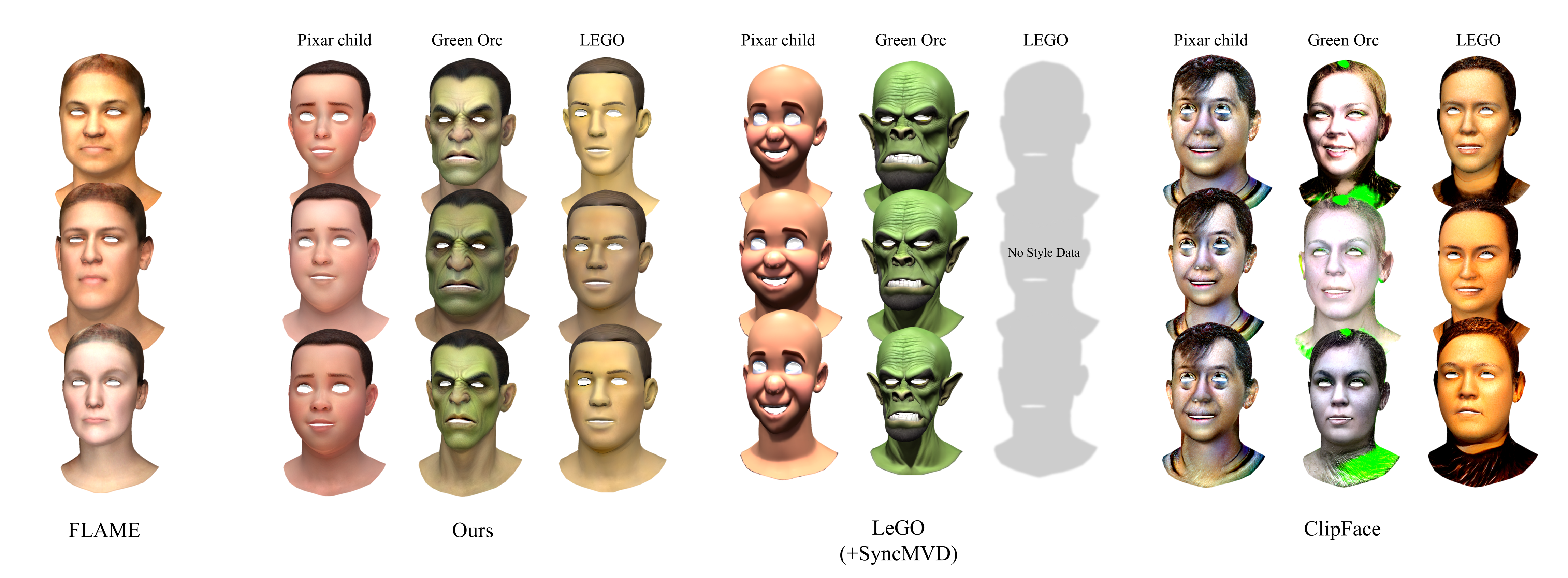}
    \vspace{-4mm}
    \caption{Qualitative comparison of randomly sampled faces generated from the realistic 3DMM and the stylized parametric face models.}\vspace{-1mm}
    \label{fig:qual}
\end{figure*}

\subsubsection{Evaluation Setting}\label{sec:eval-setting}
\paragraph*{Baselines.}  
We compared our framework against both a conventional 3DMM, FLAME~\cite{li2017learning}, and two stylized face generation methods that produce mesh-based outputs while also preserving dense vertex-level correspondence and supporting disentangled parametric control, making them functionally comparable to 3DMMs: LeGO~\cite{yoon2024lego} and ClipFace~\cite{aneja2022clipface}. 
LeGO performs shape stylization,  while ClipFace focuses on texture stylization built upon a 3DMM geometry. Because LeGO relies on one-shot learning based on 3D supervision from paired meshes of real and stylized faces, we applied it only to a subset of styles for which such data are publicly available. Moreover, as it does not perform texture stylization, we generated a single representative texture using SyncMVD~\cite{liu2023text} and applied it uniformly across all outputs.

\begin{table*}[h]
\centering
\caption{Quantitative comparison across six different styles. FLAME yields constant values. The values for LeGO and ClipFace are partially reported because LeGO requires manually crafted mesh pairs for each style while ClipFace relies on the FLAME mesh.}
\label{table:quant}
\resizebox{\textwidth}{!}{%
\begin{tabular}{l cccccc cccccc}
\toprule
\textbf{Method} 
& \multicolumn{6}{c}{\textbf{Face Diversity $\uparrow$}}
& \multicolumn{6}{c}{\textbf{Style Score $\uparrow$}} \\
\cmidrule(lr){2-7} \cmidrule(lr){8-13}   
FLAME~\cite{li2017learning}
& \multicolumn{6}{c}{12.211}
& \multicolumn{6}{c}{--} \\
\midrule 
\textbf{Style} 
& disney & pixar child & orc & baby & statue & neanderthal
& disney & pixar child & orc & baby & statue & neanderthal \\
\cmidrule(lr){2-7} \cmidrule(lr){8-13}   
LeGO~\cite{yoon2024lego}
& 10.498 & 9.836 & 9.836 & -- & -- & -- 
& 0.266 & 0.275 & 0.333 & -- & -- & -- \\
ClipFace~\cite{aneja2022clipface}
& -- & -- & -- & -- & -- & --
& 0.278 & 0.284 & 0.298 & 0.247 & \textbf{0.285} & 0.234 \\
Ours      
& 11.678 & 12.070 & 11.940 & 12.004 & 10.686 & 11.906
& \textbf{0.285} & \textbf{0.293} & \textbf{0.333} & \textbf{0.263} & 0.283 & \textbf{0.246} \\
\bottomrule 
\end{tabular}
} \vspace{-2mm}
\end{table*}

\paragraph*{Evaluation Metrics.}  
We evaluated stylized 3DMMs using the same set of 1,000 face meshes randomly sampled by varying parameters per style. Two metrics were computed {over generated meshes and corresponding renderings}:\vspace{-1.5mm}
\begin{itemize}
    \item \textbf{Face Diversity}: Measures geometric variation among the sample faces by applying PCA to vertex positions. The cumulative explained variance of the top 100 principal components is used as the metric. A higher value indicates a greater capacity to represent identity-level diversity in facial geometry.
    \item \textbf{Style Score}: Computes the average CLIP image-text similarity between a rendered image and a style prompt. A higher score reflects a stronger alignment to the input style.
\end{itemize}

\subsubsection{Image Stylization} \label{sec:image-stylization}
We first compared our stylization method with baselines, whose goal is to generate stylized faces while preserving attributes for future geometric training. The baselines consist of ControlNet~\cite{zhang2023adding}, InstructPix2Pix~\cite{li2023instruct}, PhotoMaker~\cite{li2024photomaker}, and InstantID~\cite{wang2024instantid}. As presented in Figure~\ref{fig:i2i_comparison}, ControlNet and InstructPix2Pix performed stylization following the prompt but ControlNet showed misalignment in head rotation and InstructPix2Pix removed the shoulder for the Disney character, both of which are inappropriate for deformation targets. InstantID changed the overall color of the images and failed to perform stylization correctly, following the prompt. Photomaker was unable to generate aligned images. In contrast, EAS generated stylized characters successfully following the prompt while preserving identity, expression, and alignment of the source image.

\subsubsection{Stylized 3D Morphable Face Model}  \label{sec:quantitative}

Figure~\ref{fig:qual} presents a visual comparison of the results produced by our method and baselines in three representative styles: \textit{Pixar child}, \textit{Green Orc}, and \textit{LEGO}. While FLAME produced realistic face models capable of representing diverse identities by changing shape and appearance parameters, the appearances are limited to the realistic human face domain.

Because LeGO enables expressive shape stylization through parameter-based control similar to that of $D_{style}$ of Ours, it effectively captured exaggerated facial structures. {However, because it relies on a single manually crafted pair from an artist, it shows limited diversity of the generated meshes. Furthermore, because it could not generate textures for stylized characters, the resulting meshes fail to represent textural variation within a style.
} 
ClipFace accepts arbitrary text prompts without 3D data and provides flexible texture stylization over parameter space.  
While it enables prompt-driven appearance control, we observed limitations in stylization quality:  
in some cases, the texture was overly transformed, reducing identity-level diversity; in others, the stylization was too subtle, appearing as minor artifacts over the original texture. In contrast, our method achieved identity-level face diversity comparable to FLAME, while successfully applying a wide range of text-driven styles. This enables expressive and semantically faithful stylization across a wide range of identities, without compromising structural coherence or parametric controllability.

\subsubsection{Quantitative Results}
To quantitatively evaluate our method as a stylized 3DMM, we assessed its ability to generate diverse shapes under a fixed style condition and its capacity to faithfully reflect the intended style input. As summarized in Table~\ref{table:quant}, our framework achieved the highest face diversity across most evaluated styles, indicating a stronger capability to represent identity-level shape variations under a shared stylization. Furthermore, it attained competitive or superior style scores, validating the fidelity of stylization. These results demonstrate that our model can generate expressive identity variations while maintaining consistent stylization and mesh structure—an essential feature for editable 3D avatar generation.

\subsection{Ablation Study}
\begin{figure}[t]
    \centering
    \includegraphics[width=1.02\linewidth]{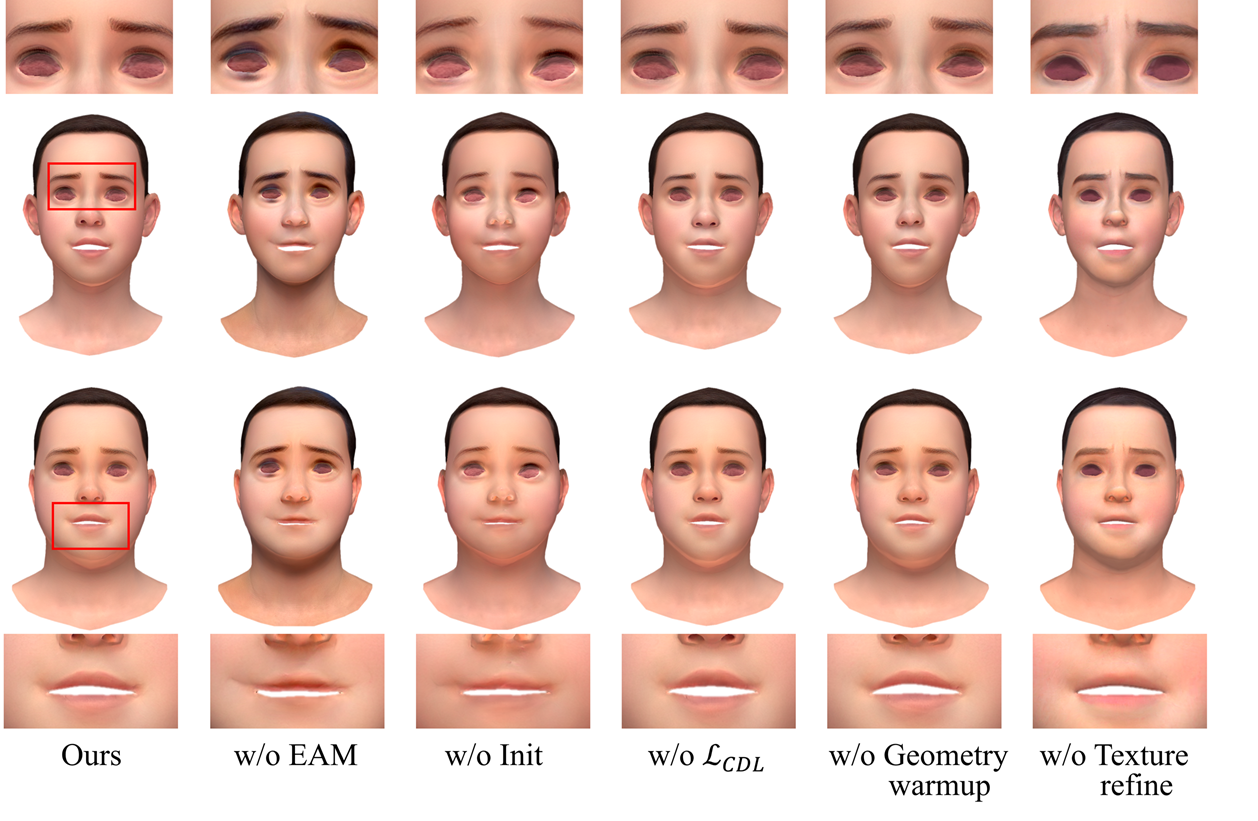} \vspace{-5mm}
    \caption{Qualitative comparison for ablation study. Each row was are rendered using the same parameters. Close-up views of the eyes and mouth are provided for visualization.}
    \label{fig:ablation} \vspace{-3mm}
\end{figure}

We conducted a series of ablation studies to investigate the contribution of each component in our framework, covering both the image stylization stage and the 3D training phase as shown in Figure~\ref{fig:ablation}. {For the ablation study, we used two different styles, Pixar child and green Orc.} We first analyzed the impact of image stylization components. Removing either EAM or latent initialization disrupted source image alignment during i2i stylization, resulting in inconsistencies between mesh geometry and texture, such as mismatched lip texture and mouth shape. Additionally, the lack of consistent style guidance degraded the overall visual quality, producing noisy textures in the eye region without EAM.

Next, we ablated the components of the 3D training pipeline. Removing the $L_{CDL}$ term led to a collapse of identity diversity, causing all shapes to converge to similar forms. This undermines the expressive range originally supported by the 3DMM, as evidenced in the quantitative results presented in Table~\ref{table:ablation}; notably, the diversity value dropped significantly for w/o $L_{CDL}$. 
In the progressive three-stage training, skipping the geometry warm-up resulted in diminished geometric deformation—evident in smaller eye shapes {and textures appear to compensate for missing geometry. 
Omitting the texture refinement phase produced textures with reduced stylistic detail} and frown-like artifacts.    These effects are reflected in the quantitative results reported in Table~\ref{table:ablation}, where both face diversity and style score dropped under ablated settings. In contrast, Ours achieved the highest scores, demonstrating the necessity of each design choice.

\begin{table}[t]
    \centering
    \caption{Ablation study on the effect of each component in terms of face diversity and style fidelity.}
    \label{table:ablation}
    \begin{tabular}{lcc}
        \toprule
        Method & Face Diversity ↑ & Style Score ↑ \\
        \midrule
        w/o EAM & 11.727 & 0.3046 \\
        w/o init & 11.819 & 0.2972 \\
        \midrule
        w/o $L_{CDL}$ & 1.812 & 0.3039 \\
        w/o warmup & 11.657 & 0.3027 \\
        w/o texture refine & - & 0.3037 \\
        \midrule
        ours & \textbf{12.005} & \textbf{0.3136} \\
        \bottomrule
    \end{tabular}
\end{table}

\begin{figure}
    \centering
    \includegraphics[width=\linewidth]{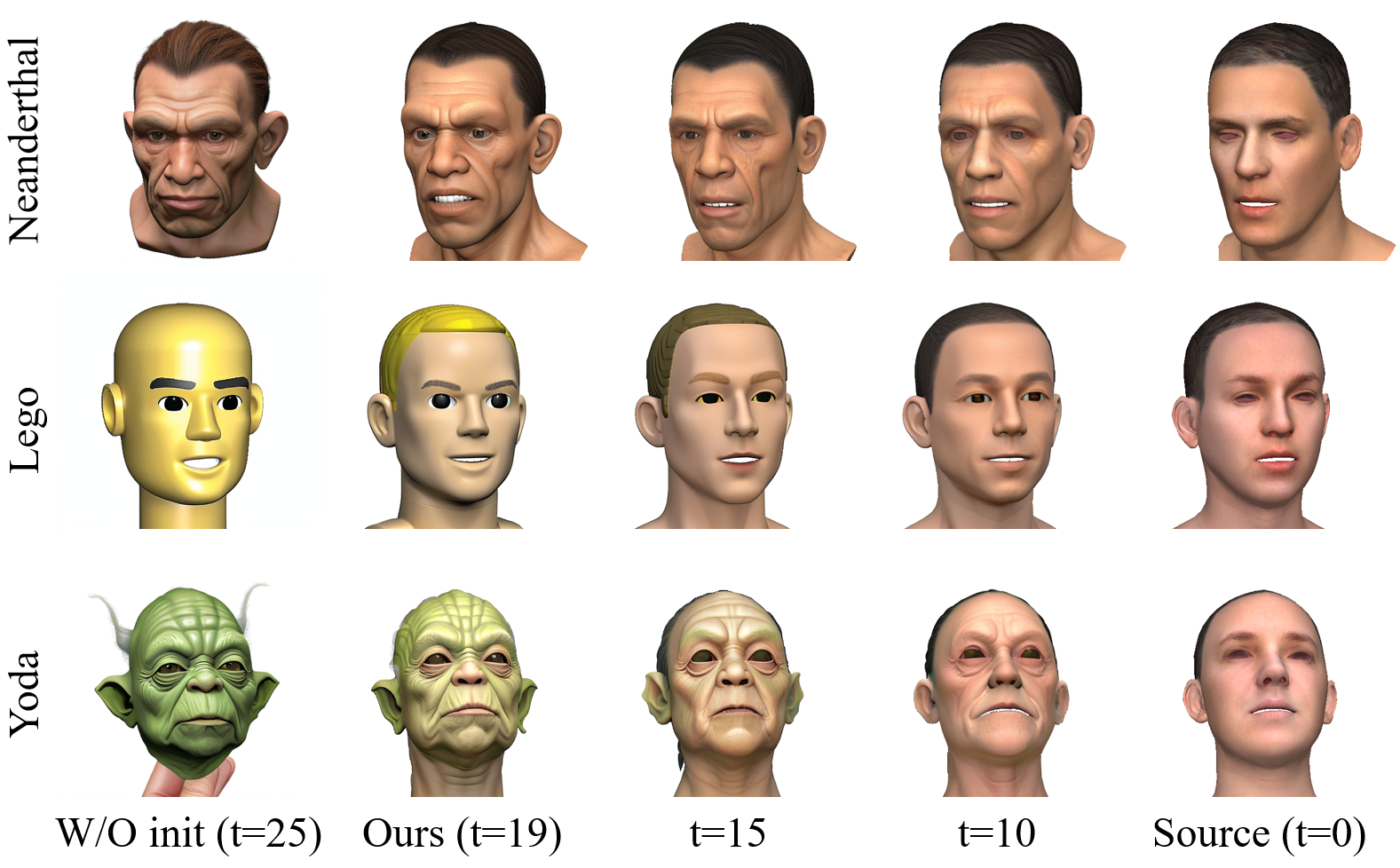}\vspace{-2mm}
    \caption{Ablation study on adjusted initialization for EAS.}
    \label{fig:init_ablation}\vspace{-2mm}
\end{figure}

We further examined the effect of our adjusted initialization strategy. As described in Section \ref{sec:stylization}, instead of sampling from pure noise, we initialized the latent variable with $x_t$ to both accelerate inference and better preserve the source’s structure and identity. Figure \ref{fig:init_ablation} compares the results: the model without using adjusted initialization achieved the strongest stylization but failed to maintain the source identity, and geometry shifted as observed in the neck region although overall facial alignment was enforced by EAM. Conversely, when using a lower value of $t$, alignment was preserved and inference became faster—because fewer diffusion steps were needed, but stylization fidelity declined sharply. The reason for the decline was that, with fewer steps (i.e., smaller $t$), EAS could not fully transform the source into the target style.

\subsection{User Study}
\begin{figure}
    \centering\vspace{-3mm}
    \includegraphics[width=\linewidth]{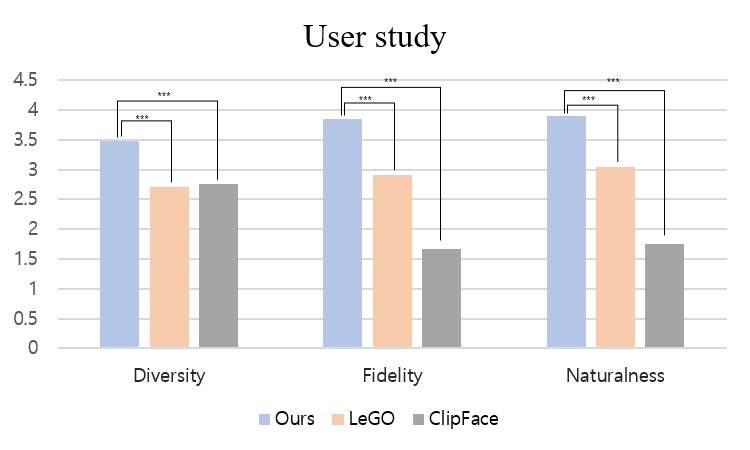}\vspace{-3mm}
    \caption{User study results. Ours showed superior results compared to baselines in Diversity, Fidelity, and Naturalness.}\vspace{-3mm}
    \label{fig:user_study}
\end{figure}

We conducted a user study to evaluate the geometric and textural diversity, style fidelity to the provided text, and overall quality of the face models generated by our method compared to LEGO+SyncMVD and CLIPFace. The study employed a 5-point Likert scale to assess three distinct styles—Disney character, Pixar child, and Green Orc—each represented by five sample images. To validate diversity, each sample image included eight randomly sampled face models from each method. A total of 21 participants (13 males, 8 females), with an average age of 28.4 years, took part in the study, providing perceptual ratings across all styles and images for each method. We conducted a one-way ANOVA to analyze statistical significance (p < 0.001). Subsequently, we performed Tukey’s HSD post hoc test, which revealed that all comparisons of our method versus each competitor were statistically significant (p < 0.001). The results, summarized in Figure~\ref{fig:user_study}, demonstrate that our approach consistently outperforms both baselines in diversity, style fidelity, and perceived quality. These findings indicate that StyleMM offers superior performance relative to both alternative methods.

\section{Applications}\vspace{-1mm}
\begin{figure}
    \centering
    \includegraphics[width=1\linewidth]{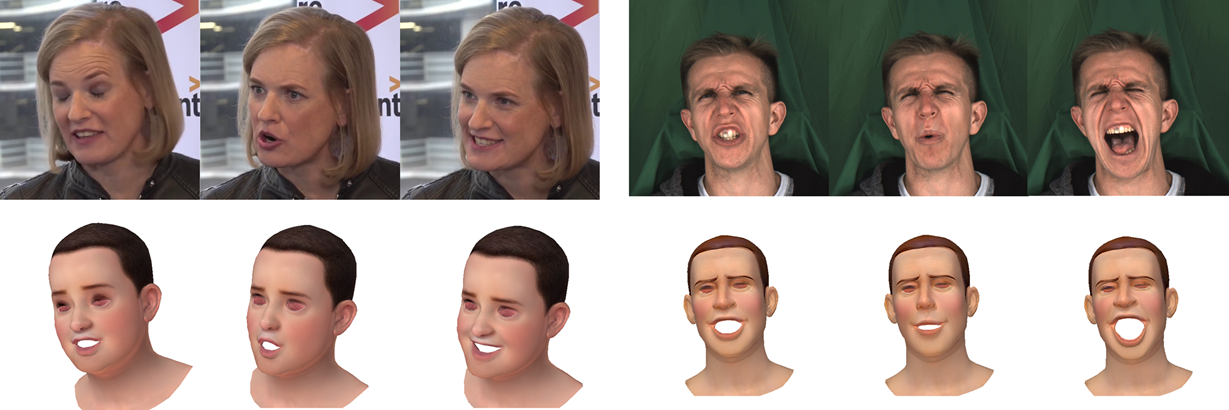}\vspace{-2mm}
    \caption{Video-driven animation results of StyleMM.}
    \label{fig:videodriven}
\end{figure}
\subsection{Video-Driven Facial Animation}
Driving animation using videos offers a practical way to create and control 3D characters, which is particularly valuable for film, game, and virtual production pipelines. 
By preserving the animatable structure of realistic 3DMM, StyleMM allows direct application of expression and pose 3DMM parameters obtained from video-based facial tracking systems.
To perform video-driven facial animation transfer, we used off-the-shelf FLAME tracking~\cite{zielonka2022towards, yan2023flameheadtracker} for the optimization of the shape, expression, and pose parameters to match the face in each video frame. 
The optimized parameters are then fed into the corresponding $D_{style}$ and $G_{style}$ of the stylized 3DMM. 
As shown in Figure~\ref{fig:videodriven}, the stylized avatar faithfully reproduces the performer's dynamics without requiring additional rigging or manual retargeting. 
This seamless retargeting enables artists and developers to leverage off-the-shelf facial capture tools for real-time or offline generation of high-quality stylized animations, significantly reducing manual workload and accelerating creative workflows.

\begin{figure}[t]
    \centering
    \vspace{-1mm}
    \includegraphics[width=\linewidth]{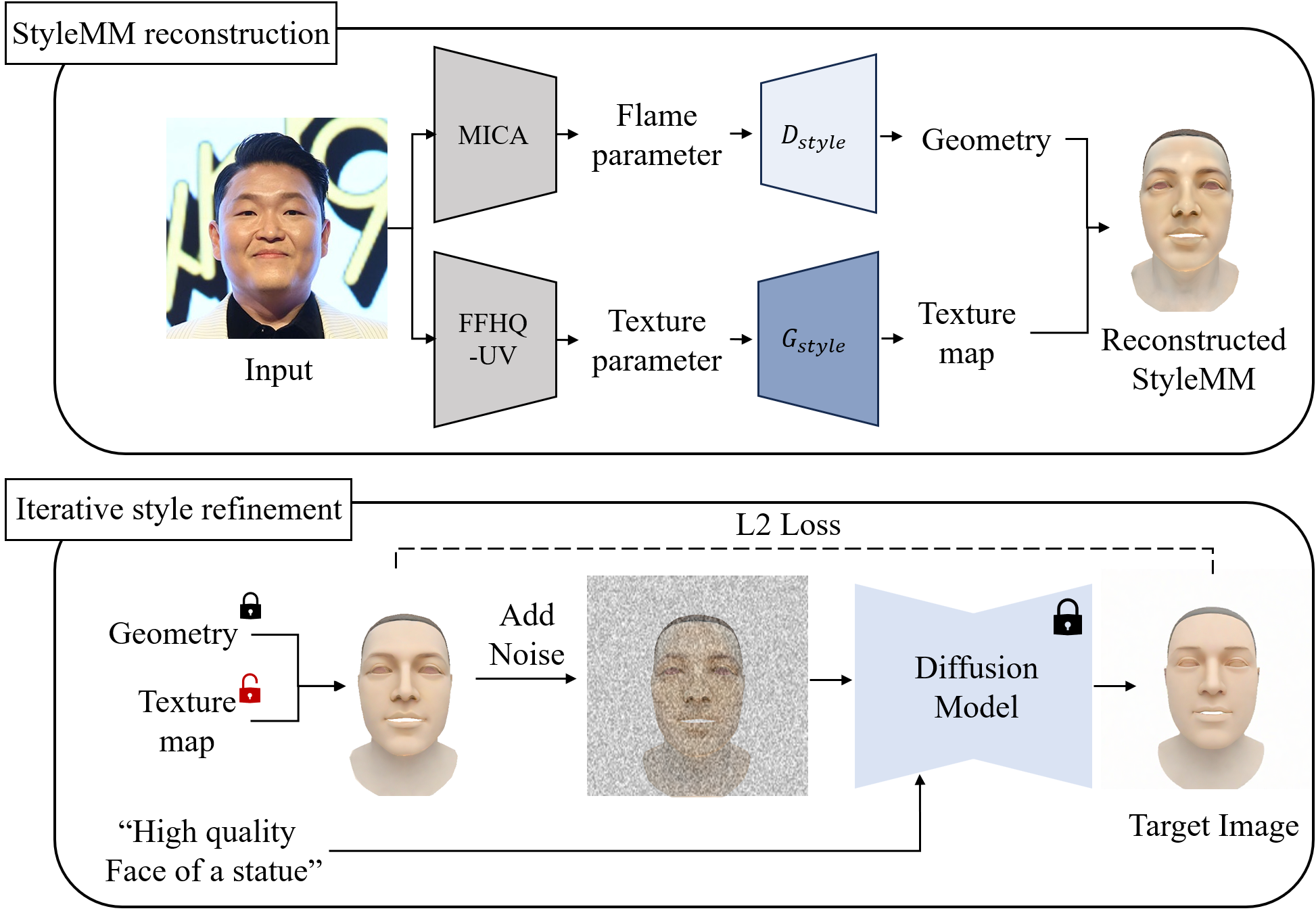}
    \caption{Two-staged stylization pipeline.}\vspace{-1mm}
    \label{fig:stylize_method}
\end{figure}

\begin{figure}[t]
    \centering
    \includegraphics[width=\linewidth]{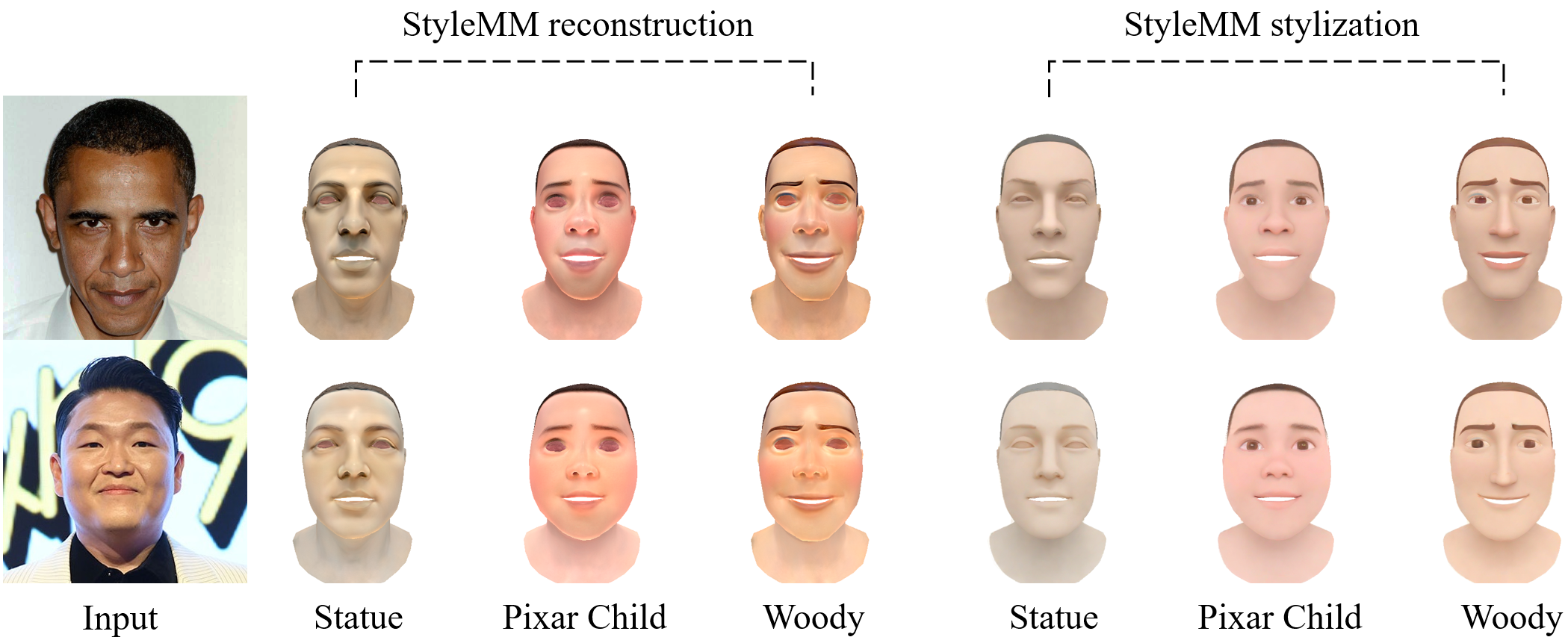}
    \vspace{-2mm}
    \caption{Stylization results from input images. When an input image is provided, the reconstruction stage generates a coarse face model (middle), and the refinement stage is applied to produce the final stylized 3D model (right). }\vspace{-3mm}
    \label{fig:stylize_result}
\end{figure}

\subsection{3D Face Stylization}\label{app:stylization}

While our primary focus is on generating stylized 3DMMs, our method naturally extends to 3D face stylization. Given an input image, we first estimate shape, expression, and texture parameters using off-the-shelf methods such as MICA~\cite{zielonka2022towards} and FFHQ-UV~\cite{bai2023ffhq}. These parameters are then fed into our stylized 3DMM networks, $D_{style}$ and $G_{style}$, to reconstruct geometry and appearance that adhere to both the original identity and the desired target style. This process is presented in the first row of Figure~\ref{fig:stylize_method}. Although this StyleMM reconstruction already produces a stylized 3D face, it may lack the smoothness typically expected in high-quality face stylizations. To address these issues, we introduce a subsequent texture refinement stage.

Following the approach of i2i translation  studies~\cite{wu2024portrait3d,meng2021sdedit,rombach2022high}, our texture refinement process operates directly on the reconstructed UV map. Because the source image has already been stylized by StyleMM, we only inject a small amount of noise and perform a denoising step. We use this denoised image as a target stylized face and update the texture map iteratively. Whereas our EAS initialization adds noise over 76\% of the total diffusion steps; this iterative refinement uses just 25\% of the diffusion budget and runs for 400 optimization iterations. We employ a 4-step SDXL-Turbo diffusion model~\cite{podell2023sdxl,sauer2024adversarial} to ensure rapid convergence. As demonstrated in Figure~\ref{fig:stylize_result}, this refinement produces a smoother, higher-fidelity texture while preserving the stylized geometry obtained from the StyleMM networks.

\subsection{Adding Facial Attribute}\label{app:attribute}
\begin{figure}\vspace{-1mm}
    \centering
    \includegraphics[width=1\linewidth]{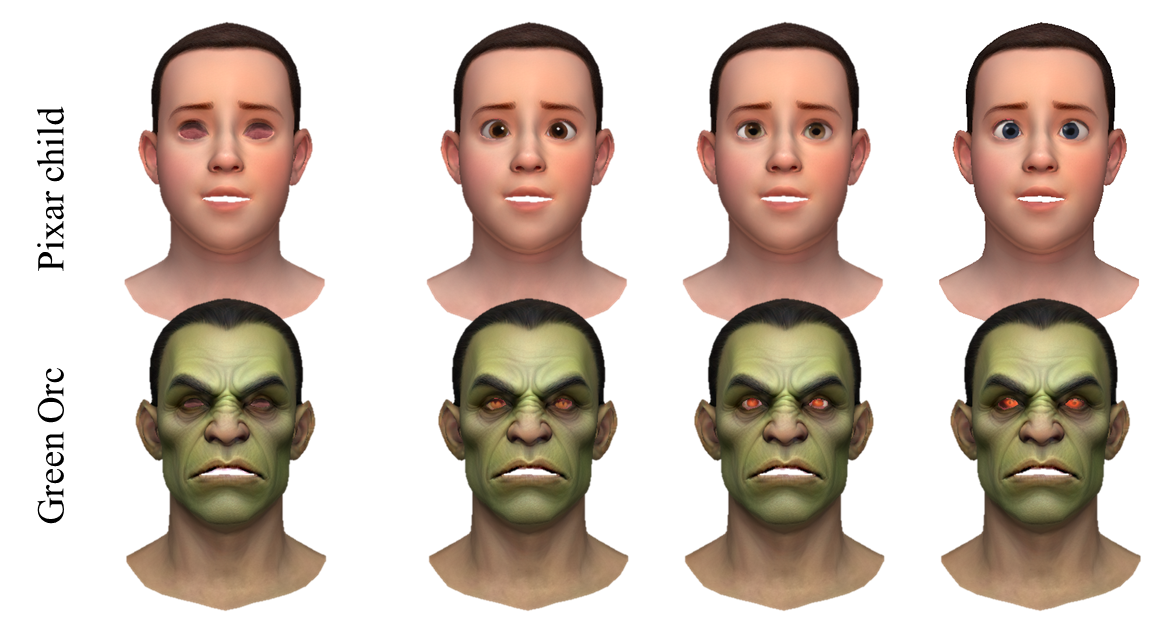}\vspace{-2mm}
    \caption{Left: original StyleMM output; Right: eyeball post-processing results.}\vspace{-2mm}
    \label{fig:app_eye}
\end{figure}

In our original StyleMM pipeline, we excluded the eyeball geometry during training, following the deformation-network strategy of LeGO~\cite{yoon2024lego}. To restore facial details omitted from StyleMM’s parametric output, we performed post-processing by adding eyeballs. Using the eye socket center vertex and its distance to the surrounding eyelid vertices, we inserted the original FLAME rigid eyeball meshes and scaled them accordingly. For texture optimization, we utilized EAS stylization to generate references for the L2 loss applied to the rendered outputs. We then added a total variation loss to the texture map to promote spatially uniform optimization across the surface. During this optimization, all other geometry, texture, and network parameters were fixed.
As shown in Figure~\ref{fig:app_eye}, this approach produced realistic eyeballs in the StyleMM outputs. To evaluate the effect of this attribute addition, we conducted a user study with 15 participants, who compared results with and without eyeballs in terms of style fidelity to the textual prompt and overall naturalness. Using Pixar child and green orc models in an A/B test, the eyeball-enhanced versions were strongly preferred, achieving 96.67\% in style fidelity and 91.33\% in naturalness.



\section{Conclusion}\vspace{-1mm}
In this work, we have introduced StyleMM, a novel framework for constructing stylized 3DMMs directly from text-driven image translations. By defining the three core requirements of stylized 3DMMs—\emph{maintained correspondence}, \emph{disentangled control}, and \emph{stylization beyond realistic geometry and texture}—we have positioned stylized 3DMMs as a distinct research paradigm, particularly important for applications in film, animation, and game production. Unlike prior face stylization methods, StyleMM simultaneously satisfies all three criteria by leveraging text-guided i2i translations to fine-tune both a surface deformation network and a texture generator originally designed for realistic 3DMMs.

Our approach makes stylized 3DMMs accessible to users without requiring a stylized 3D dataset by utilizing EAS, a novel image stylization method. Through the three-stage training pipeline and CDL, StyleMM maintains mesh correspondence across diverse identities, ensures independent control over geometry and texture, and achieves a wide spectrum of stylistic variations in both geometry and appearance. Quantitative and qualitative evaluations demonstrate that StyleMM outperforms existing baselines by preserving identity-level variation while extending beyond realistic models into highly expressive, stylized domains.

\begin{figure}[t]
    \centering
    \includegraphics[width=0.9\linewidth]{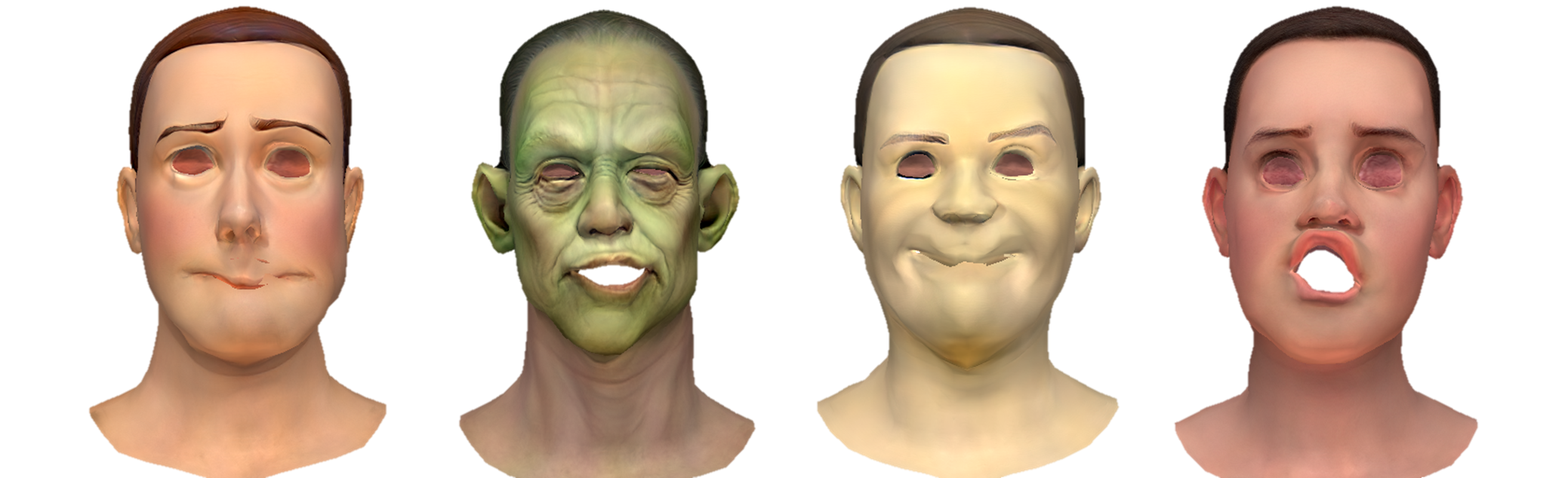}\vspace{-2mm}
    \caption{Self-intersection under exaggerated expressions}
    \label{fig:intersection}\vspace{-4mm}
\end{figure}

\subsection{Limitations and Future Work}
A primary limitation of StyleMM lies in the trade-off between stylization capability and geometric consistency. In EAS, applying extreme stylization can introduce geometry and identity misalignments in the i2i-generated outputs, as shown by the W/O init results in Figure~\ref{fig:init_ablation}. This can result in unexpected mesh deformations. Although EAS mitigates misalignment of facial landmarks and expressions during stylization, residual discrepancies may still impair dense correspondence under highly artistic style transfers.
To reduce misalignment for extreme artistic styles, future work could investigate improved alignment strategies that dynamically adjust landmark weighting during stylization. Integrating a confidence-weighted attribute preservation module could more effectively reconcile aggressive geometric deviations with consistent mesh structure. 

In 3DMM fine-tuning, strong mesh stabilization losses are applied to maintain plausible geometry, but they can also suppress sharp stylistic details such as pointed ears and fine wrinkles, reducing intended geometric expressiveness.
Incorporating multi-scale structural priors—such as Laplacian regularization or mesh spectral embeddings—into the CDL may better preserve both global and local geometry under strong style shifts, enabling more radical stylizations without compromising correspondence. Additionally, integrating displacement mapping could further enhance geometry by disentangling it from texture, although this may necessitate additional data for parameterization.

Another limitation is self-intersection under exaggerated expressions. While our method generally produces meshes of superior quality compared to previous studies due to the adoption of a pre-trained surface deformation network, self-intersection can still occur with exaggerated expression as shown in Figure~\ref{fig:intersection}. One possible solution might be to incorporate more diverse expressions during training.

Pursuing these directions may help close the existing gaps between unconstrained text-driven stylization and strict geometric coherence, paving the way for stylized 3DMMs to become standard tools in animation, virtual production, and avatar creation not only for large groups of professional artists but also for small studios and game developers.

\vspace{-2mm}
\section*{Acknowledgements}\vspace{-2mm}
{This work was supported by Institute of Information \& Communications Technology Planning \& Evaluation(IITP) grant funded by the Korea government(MSIT) (No.RS-2024-00439499, Generating Hyper-Realistic to Extremely-stylized Face Avatar with Varied Speech Speed and Context-based Emotional Expression)}\vspace{-2mm}
\bibliographystyle{eg-alpha-doi} 
\bibliography{egbibsample}

\end{document}